\documentclass[12pt,pre,eqsecnum,preprint,amsfonts]{revtex4}

\usepackage{epsf}

\newcommand{\beq}{\begin{equation}}
\newcommand{\eeq}{\end{equation}}
\newcommand{\beqs}{\begin{eqnarray}}
\newcommand{\eeqs}{\end{eqnarray}}

\newtheorem{cor}{Corollary}[section]
\newtheorem{lem}{Lemma}[section]

\newtheorem{propo}{Proposition}[section]
\newtheorem{defi}{Definition}[section]

\begin{document}

\title{Asymptotic enumeration of independent sets on the Sierpinski gasket}

\author{Shu-Chiuan Chang$^{a,b}$} 
\email{scchang@mail.ncku.edu.tw} 

\affiliation{(a) \ Department of Physics \\
National Cheng Kung University \\
Tainan 70101, Taiwan} 

\affiliation{(b) \ Physics Division \\
National Center for Theoretical Science \\
National Taiwan University \\
Taipei 10617, Taiwan} 

\author{Lung-Chi Chen$^{c}$} 
\email{lcchen@math.fju.edu.tw}

\affiliation{(c) \ Department of Mathematics \\
Fu Jen Catholic University \\
Taipei 24205, Taiwan }

\author{Weigen Yan$^{d}$} 
\email{weigenyan@263.net}

\affiliation{(d) \ School of Sciences \\
Jimei University \\
Xiamen 361021, China }

\begin{abstract}

The number of independent sets is equivalent to the partition function of the hard-core lattice gas model with nearest-neighbor exclusion and unit activity. We study the number of independent sets $m_{d,b}(n)$ on the generalized Sierpinski gasket $SG_{d,b}(n)$ at stage $n$ with dimension $d$ equal to two, three and four for $b=2$, and layer $b$ equal to three for $d=2$. The upper and lower bounds for the asymptotic growth constant, defined as $z_{SG_{d,b}}=\lim_{v \to \infty} \ln m_{d,b}(n)/v$ where $v$ is the number of vertices, on these Sierpinski gaskets are derived in terms of the results at a certain stage. The numerical values of these $z_{SG_{d,b}}$ are evaluated with more than a hundred significant figures accurate. We also conjecture the upper and lower bounds for the asymptotic growth constant $z_{SG_{d,2}}$ with general $d$.

\end{abstract}

\maketitle

\pagestyle{plain}
\pagenumbering{arabic}

\section{Introduction}
\label{introduction}

The lattice gas with repulsive pair interaction is an important model in statistical mechanics \cite{runnels,brightwell,heringa,guo}. For the special case with hard-core nearest-neighbor exclusion such that each site can be occupied by at most one particle and no pair of adjacent sites can be simultaneously occupied, the partition function of the lattice gas coincides with the independence polynomial in combinatorics \cite{gutman,scott}. This model is a problem of interest in mathematics \cite{berg,haggstrom,kahn,dyer}. While an activity (or fugacity) $\lambda$ can be associated to each occupied site, the special case with $\lambda=1$ counts the number of independent (vertex) sets $N_{IS}(G)$ on a graph $G$ \cite{prodinger}. Kaplansky considered the number of $k$-element independent sets on the path and circuit graphs almost 70 years ago \cite{kaplansky}. For a graph $G$ with $v(G)$ vertices, the number of independent sets grows exponentially when $v(G)$ is large. For the $m \times n$ grid graph, i.e. the square lattice (sq), it was shown that the limit $\lim_{m,n \to \infty} N_{IS}(sq)^{1/mn}$ exists and its upper and lower bounds were estimated \cite{calkin}. The number of independent sets and its bounds had been considered on various graphs \cite{linek, law, zhao}.  

It is of interest to consider independent sets on self-similar fractal lattices which have scaling invariance rather than translational invariance \cite{teufl}. Fractals are geometric structures of (generally noninteger) Hausdorff dimension realized by repeated construction of an elementary shape on progressively smaller length scales \cite{mandelbrot,Falconer}. A well-known example of a fractal is the Sierpinski gasket which has been extensively studied in several contexts \cite{Gefen80, Gefen81, Rammal, Alexander, Domany, Gefen8384, Guyer, Kusuoka, Dhar97, Daerden, Dhar05, sts, sfs, css, ds, dms, hs}. 

We shall derive the recursion relations for the numbers of independent sets on the Sierpinski gasket with dimension equal to two, three and four, and determine the asymptotic growth constants. We shall also consider the number of independent sets on the generalized two-dimensional Sierpinski gasket with layer equal to three.

\section{Preliminaries}
\label{preliminary}

We first recall some relevant definitions for graphs and the Sierpinski gasket in this section. A connected graph (without loops) $G=(V,E)$ is defined by its vertex (site) and edge (bond) sets $V$ and $E$ \cite{fh,bbook}.  Let $v(G)=|V|$ be the number of vertices and $e(G)=|E|$ the number of edges in $G$.  The degree or coordination number $k_i$ of a vertex $v_i \in V$ is the number of edges attached to it.  A $k$-regular graph is a graph with the property that each of its vertices has the same degree $k$. An independent set is a subset of the vertices such that any two of them are not adjacent.

When the number of independent sets $N_{IS}(G)$ grows exponentially with $v(G)$ as $v(G) \to \infty$, let us define a constant $z_G$ describing this exponential growth:
\beq
z_G = \lim_{v(G) \to \infty} \frac{\ln N_{IS}(G)}{v(G)} \ ,
\label{zdef}
\eeq
where $G$, when used as a subscript in this manner, implicitly refers to
the thermodynamic limit. We will see that the limit in Eq. (\ref{zdef}) exists for the Sierpinski gasket considered in this paper.

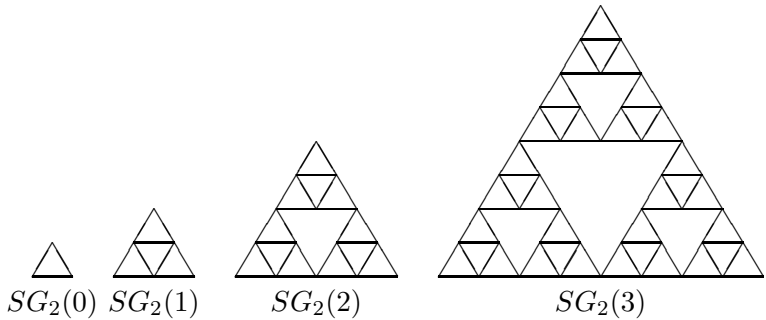
\begin{figure}[htbp]
\unitlength 0.9mm \hspace*{3mm}
\begin{picture}(108,40)
\put(0,0){\line(1,0){6}}
\put(0,0){\line(3,5){3}}
\put(6,0){\line(-3,5){3}}
\put(3,-4){\makebox(0,0){$SG_2(0)$}}
\put(12,0){\line(1,0){12}}
\put(12,0){\line(3,5){6}}
\put(24,0){\line(-3,5){6}}
\put(15,5){\line(1,0){6}}
\put(18,0){\line(3,5){3}}
\put(18,0){\line(-3,5){3}}
\put(18,-4){\makebox(0,0){$SG_2(1)$}}
\put(30,0){\line(1,0){24}}
\put(30,0){\line(3,5){12}}
\put(54,0){\line(-3,5){12}}
\put(36,10){\line(1,0){12}}
\put(42,0){\line(3,5){6}}
\put(42,0){\line(-3,5){6}}
\multiput(33,5)(12,0){2}{\line(1,0){6}}
\multiput(36,0)(12,0){2}{\line(3,5){3}}
\multiput(36,0)(12,0){2}{\line(-3,5){3}}
\put(39,15){\line(1,0){6}}
\put(42,10){\line(3,5){3}}
\put(42,10){\line(-3,5){3}}
\put(42,-4){\makebox(0,0){$SG_2(2)$}}
\put(60,0){\line(1,0){48}}
\put(72,20){\line(1,0){24}}
\put(60,0){\line(3,5){24}}
\put(84,0){\line(3,5){12}}
\put(84,0){\line(-3,5){12}}
\put(108,0){\line(-3,5){24}}
\put(66,10){\line(1,0){12}}
\put(90,10){\line(1,0){12}}
\put(78,30){\line(1,0){12}}
\put(72,0){\line(3,5){6}}
\put(96,0){\line(3,5){6}}
\put(84,20){\line(3,5){6}}
\put(72,0){\line(-3,5){6}}
\put(96,0){\line(-3,5){6}}
\put(84,20){\line(-3,5){6}}
\multiput(63,5)(12,0){4}{\line(1,0){6}}
\multiput(66,0)(12,0){4}{\line(3,5){3}}
\multiput(66,0)(12,0){4}{\line(-3,5){3}}
\multiput(69,15)(24,0){2}{\line(1,0){6}}
\multiput(72,10)(24,0){2}{\line(3,5){3}}
\multiput(72,10)(24,0){2}{\line(-3,5){3}}
\multiput(75,25)(12,0){2}{\line(1,0){6}}
\multiput(78,20)(12,0){2}{\line(3,5){3}}
\multiput(78,20)(12,0){2}{\line(-3,5){3}}
\put(81,35){\line(1,0){6}}
\put(84,30){\line(3,5){3}}
\put(84,30){\line(-3,5){3}}
\put(84,-4){\makebox(0,0){$SG_2(3)$}}
\end{picture}

\vspace*{5mm}
\caption{\footnotesize{The first four stages $n=0,1,2,3$ of the two-dimensional Sierpinski gasket $SG_2(n)$.}} 
\label{sgfig}
\end{figure}

The construction of the two-dimensional Sierpinski gasket $SG_2(n)$ at stage $n$ is shown in Fig. \ref{sgfig}. At stage $n=0$, it is an equilateral triangle; while stage $(n+1)$ is obtained by the juxtaposition of three $n$-stage structures. In general, the Sierpinski gaskets $SG_d$ can be built in any Euclidean dimension $d$ with fractal dimension $D=\ln(d+1)/\ln2$ \cite{Gefen81}. For the Sierpinski gasket $SG_d(n)$, the numbers of edges and vertices are given by 
\beq
e(SG_d(n)) = {d+1 \choose 2} (d+1)^n = \frac{d}{2} (d+1)^{n+1} \ ,
\label{e}
\eeq
\beq
v(SG_d(n)) = \frac{d+1}{2} [(d+1)^n+1] \ .
\label{v}
\eeq
Except the $(d+1)$ outmost vertices which have degree $d$, all other vertices of $SG_d(n)$ have degree $2d$. In the large $n$ limit, $SG_d$ is $2d$-regular.

\begin{figure}[htbp]
\unitlength 0.9mm \hspace*{3mm}
\begin{picture}(108,45)
\put(0,0){\line(1,0){18}}
\put(3,5){\line(1,0){12}}
\put(6,10){\line(1,0){6}}
\put(0,0){\line(3,5){9}}
\put(6,0){\line(3,5){6}}
\put(12,0){\line(3,5){3}}
\put(18,0){\line(-3,5){9}}
\put(12,0){\line(-3,5){6}}
\put(6,0){\line(-3,5){3}}
\put(9,-4){\makebox(0,0){$SG_{2,3}(1)$}}
\put(24,0){\line(1,0){54}}
\put(33,15){\line(1,0){36}}
\put(42,30){\line(1,0){18}}
\put(24,0){\line(3,5){27}}
\put(42,0){\line(3,5){18}}
\put(60,0){\line(3,5){9}}
\put(78,0){\line(-3,5){27}}
\put(60,0){\line(-3,5){18}}
\put(42,0){\line(-3,5){9}}
\multiput(27,5)(18,0){3}{\line(1,0){12}}
\multiput(30,10)(18,0){3}{\line(1,0){6}}
\multiput(30,0)(18,0){3}{\line(3,5){6}}
\multiput(36,0)(18,0){3}{\line(3,5){3}}
\multiput(36,0)(18,0){3}{\line(-3,5){6}}
\multiput(30,0)(18,0){3}{\line(-3,5){3}}
\multiput(36,20)(18,0){2}{\line(1,0){12}}
\multiput(39,25)(18,0){2}{\line(1,0){6}}
\multiput(39,15)(18,0){2}{\line(3,5){6}}
\multiput(45,15)(18,0){2}{\line(3,5){3}}
\multiput(45,15)(18,0){2}{\line(-3,5){6}}
\multiput(39,15)(18,0){2}{\line(-3,5){3}}
\put(45,35){\line(1,0){12}}
\put(48,40){\line(1,0){6}}
\put(48,30){\line(3,5){6}}
\put(54,30){\line(3,5){3}}
\put(54,30){\line(-3,5){6}}
\put(48,30){\line(-3,5){3}}
\put(48,-4){\makebox(0,0){$SG_{2,3}(2)$}}
\put(84,0){\line(1,0){24}}
\put(87,5){\line(1,0){18}}
\put(90,10){\line(1,0){12}}
\put(93,15){\line(1,0){6}}
\put(84,0){\line(3,5){12}}
\put(90,0){\line(3,5){9}}
\put(96,0){\line(3,5){6}}
\put(102,0){\line(3,5){3}}
\put(108,0){\line(-3,5){12}}
\put(102,0){\line(-3,5){9}}
\put(96,0){\line(-3,5){6}}
\put(90,0){\line(-3,5){3}}
\put(96,-4){\makebox(0,0){$SG_{2,4}(1)$}}
\end{picture}

\vspace*{5mm}
\caption{\footnotesize{The generalized two-dimensional Sierpinski gasket $SG_{2,b}(n)$ with $b=3$ at stage $n=1, 2$ and $b=4$ at stage $n=1$.}} 
\label{sgbfig}
\end{figure}
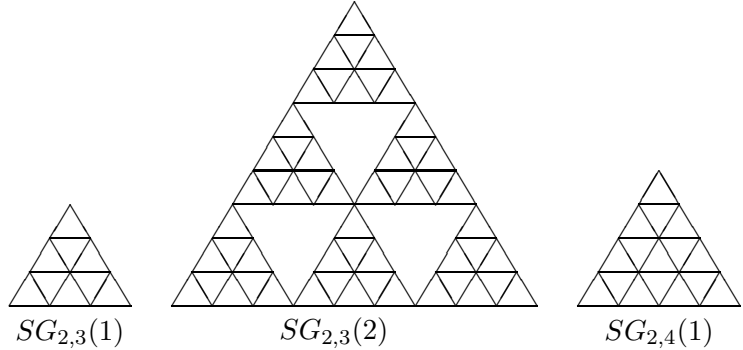

The Sierpinski gasket can be generalized, denoted as $SG_{d,b}(n)$, by introducing the side length $b$ which is an integer larger or equal to two \cite{Hilfer}. The generalized Sierpinski gasket at stage $n+1$ is constructed from $b$ layers of stage $n$ hypertetrahedrons. The two-dimensional $SG_{2,b}(n)$ with $b=3$ at stage $n=1, 2$ and $b=4$ at stage $n=1$ are illustrated in Fig. \ref{sgbfig}. The ordinary Sierpinski gasket $SG_d(n)$ corresponds to the $b=2$ case, where the index $b$ is neglected for simplicity. The Hausdorff dimension for $SG_{d,b}$ is given by $D=\ln {b+d-1 \choose d} / \ln b$ \cite{Hilfer}. Notice that $SG_{d,b}$ is not $k$-regular even in the thermodynamic limit.

\section{The number of independent sets on $SG_2(n)$}
\label{sectionIII}

In this section we derive the asymptotic growth constant for the number of independent sets on the two-dimensional Sierpinski gasket $SG_2(n)$ in detail. Let us start with the definitions of the quantities to be used.

\bigskip

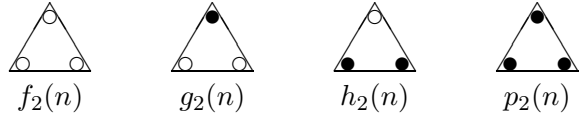
\begin{figure}[htbp]
\unitlength 1.8mm 
\begin{picture}(42,5)
\put(0,0){\line(1,0){6}}
\put(0,0){\line(3,5){3}}
\put(6,0){\line(-3,5){3}}
\multiput(1,0.5)(4,0){2}{\circle{1}}
\put(3,4){\circle{1}}
\put(3,-2){\makebox(0,0){$f_2(n)$}}
\put(12,0){\line(1,0){6}}
\put(12,0){\line(3,5){3}}
\put(18,0){\line(-3,5){3}}
\put(13,0.5){\circle{1}}
\put(17,0.5){\circle{1}}
\put(15,4){\circle*{1}}
\put(15,-2){\makebox(0,0){$g_2(n)$}}
\put(24,0){\line(1,0){6}}
\put(24,0){\line(3,5){3}}
\put(30,0){\line(-3,5){3}}
\put(25,0.5){\circle*{1}}
\put(29,0.5){\circle*{1}}
\put(27,4){\circle{1}}
\put(27,-2){\makebox(0,0){$h_2(n)$}}
\put(36,0){\line(1,0){6}}
\put(36,0){\line(3,5){3}}
\put(42,0){\line(-3,5){3}}
\multiput(37,0.5)(4,0){2}{\circle*{1}}
\put(39,4){\circle*{1}}
\put(39,-2){\makebox(0,0){$p_2(n)$}}
\end{picture}

\vspace*{5mm}
\caption{\footnotesize{Illustration for the configurations $f_2(n)$, $g_2(n)$, $h_2(n)$, and $p_2(n)$. Only the three outmost vertices are shown explicitly, where a solid circle is in the vertex subset and an open circle is not.}} 
\label{fghtfig}
\end{figure}

\begin{defi} \label{defisg2} Consider the generalized two-dimensional Sierpinski gasket $SG_{2,b}(n)$ at stage $n$. (i) Define $m_{2,b}(n) \equiv N_{IS}(SG_{2,b}(n))$ as the number of independent sets. (ii) Define $f_{2,b}(n)$ as the number of independent sets such that all three outmost vertices are not in the vertex subset. (ii) Define $g_{2,b}(n)$ as the number of independent sets such that one certain outmost vertex, say the topmost vertex as illustrated in Fig. \ref{fghtfig} for ordinary Sierpinski gasket, is in the vertex subset. (iii) Define $h_{2,b}(n)$ as the number of independent sets such that two certain outmost vertices, say the leftmost and rightmost vertices as illustrated in Fig. \ref{fghtfig}, are in the vertex subset. (iv) Define $p_{2,b}(n)$ as the number of independent sets such that all three outmost vertices are in the vertex subset.
\end{defi}

\bigskip

Since we only consider the ordinary Sierpinski gasket in this section, we use the notations $m_2(n)$, $f_2(n)$, $g_2(n)$, $h_2(n)$ and $p_2(n)$ for simplicity. They are illustrated in Fig. \ref{fghtfig}, where only the outmost vertices are shown. Because of rotational symmetry, there are three possible $g_2(n)$ and three possible $h_2(n)$ such that
\beq
m_2(n) = f_2(n)+3g_2(n)+3h_2(n)+p_2(n) \ .
\label{meq}
\eeq
The initial values at stage zero are $f_2(0)=1$, $g_2(0)=1$, $h_2(0)=0$, $p_2(0)=0$ and $m_2(0)=4$. The purpose of this section is to obtain the asymptotic behavior of $m_2(n)$ as follows.
The four quantities $f_2(n)$, $g_2(n)$, $h_2(n)$ and $p_2(n)$ satisfy recursion relations.

\begin{lem} \label{lemmasg2r} For any non-negative integer $n$,
\beq
f_2(n+1) = f_2^3(n) + 3f_2(n)g_2^2(n) + 3g_2^2(n)h_2(n) + h_2^3(n) \ , 
\label{feq}
\eeq
\beqs
\lefteqn{g_2(n+1)} \cr & = & f_2^2(n)g_2(n) + 2f_2(n)g_2(n)h_2(n) + g_2^3(n) + 2g_2(n)h_2^2(n) + g_2^2(n)p_2(n) + h_2^2(n)p_2(n) \ , 
\label{geq}
\eeqs
\beqs
\lefteqn{h_2(n+1)} \cr & = & f_2(n)g_2^2(n) + f_2(n)h_2^2(n) + 2g_2^2(n)h_2(n) + h_2^3(n) + 2g_2(n)h_2(n)p_2(n) + h_2(n)p_2^2(n) \ ,
\label{heq}
\eeqs
\beq
p_2(n+1) = g_2^3(n) + 3g_2(n)h_2^2(n) + 3h_2^2(n)p_2(n) + p_2^3(n) \ .
\label{peq}
\eeq
\end{lem}

\begin{figure}[htbp]
\unitlength 1.2mm 
\begin{picture}(90,12)
\put(0,0){\line(1,0){12}}
\put(0,0){\line(3,5){6}}
\put(12,0){\line(-3,5){6}}
\multiput(1,0.5)(10,0){2}{\circle{1}}
\put(6,9){\circle{1}}
\put(15,5){\makebox(0,0){$=$}}
\put(18,0){\line(1,0){12}}
\put(18,0){\line(3,5){6}}
\put(24,0){\line(3,5){3}}
\put(24,0){\line(-3,5){3}}
\put(21,5){\line(1,0){6}}
\put(30,0){\line(-3,5){6}}
\put(19,0.5){\circle{1}}
\put(21,4){\circle{1}}
\put(23,0.5){\circle{1}}
\put(22,5.5){\circle{1}}
\put(24,9){\circle{1}}
\put(26,5.5){\circle{1}}
\put(25,0.5){\circle{1}}
\put(27,4){\circle{1}}
\put(29,0.5){\circle{1}}
\put(33,5){\makebox(0,0){$+$}}
\put(36,0){\line(1,0){12}}
\put(36,0){\line(3,5){6}}
\put(42,0){\line(3,5){3}}
\put(42,0){\line(-3,5){3}}
\put(39,5){\line(1,0){6}}
\put(48,0){\line(-3,5){6}}
\put(37,0.5){\circle{1}}
\put(39,4){\circle{1}}
\put(41,0.5){\circle*{1}}
\put(40,5.5){\circle{1}}
\put(42,9){\circle{1}}
\put(44,5.5){\circle{1}}
\put(43,0.5){\circle*{1}}
\put(45,4){\circle{1}}
\put(47,0.5){\circle{1}}
\put(49,5){\makebox(0,0){$\times 3$}}
\put(54,5){\makebox(0,0){$+$}}
\put(57,0){\line(1,0){12}}
\put(57,0){\line(3,5){6}}
\put(63,0){\line(3,5){3}}
\put(63,0){\line(-3,5){3}}
\put(60,5){\line(1,0){6}}
\put(69,0){\line(-3,5){6}}
\put(58,0.5){\circle{1}}
\put(60,4){\circle*{1}}
\put(62,0.5){\circle{1}}
\put(61,5.5){\circle*{1}}
\put(63,9){\circle{1}}
\put(65,5.5){\circle*{1}}
\put(64,0.5){\circle{1}}
\put(66,4){\circle*{1}}
\put(68,0.5){\circle{1}}
\put(70,5){\makebox(0,0){$\times 3$}}
\put(75,5){\makebox(0,0){$+$}}
\put(78,0){\line(1,0){12}}
\put(78,0){\line(3,5){6}}
\put(84,0){\line(3,5){3}}
\put(84,0){\line(-3,5){3}}
\put(81,5){\line(1,0){6}}
\put(90,0){\line(-3,5){6}}
\put(79,0.5){\circle{1}}
\put(81,4){\circle*{1}}
\put(83,0.5){\circle*{1}}
\put(82,5.5){\circle*{1}}
\put(84,9){\circle{1}}
\put(86,5.5){\circle*{1}}
\put(85,0.5){\circle*{1}}
\put(87,4){\circle*{1}}
\put(89,0.5){\circle{1}}
\end{picture}

\caption{\footnotesize{Illustration for the expression of  $f_2(n+1)$. The multiplication of three on the right-hand-side corresponds to the three possible orientations of $SG_2(n+1)$.}} 
\label{ffig}
\end{figure}
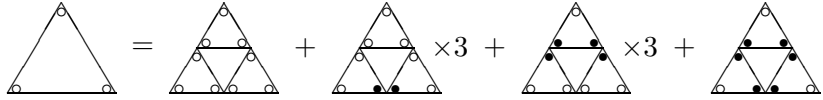

{\sl Proof} \quad 
The Sierpinski gasket $SG_2(n+1)$ is composed of three $SG_2(n)$ with three pairs of vertices identified. The number $f_2(n+1)$ consists of (i) one configuration where all three $SG_2(n)$ belong to the class that is enumerated by $f_2(n)$; (ii) three configurations where one of the $SG_2(n)$ belongs to the class enumerated by $f_2(n)$ and the other two belong to the class enumerated by $g_2(n)$; (iii) three configurations where two of the $SG_2(n)$ belong to the class enumerated by $g_2(n)$ and the other one belongs to the class enumerated by $h_2(n)$; (iv) one configuration where all three $SG_2(n)$ belongs to the class enumerated by $h_2(n)$ as illustrated in Fig. \ref{ffig}. Eq. (\ref{feq}) is verified by adding these configurations.

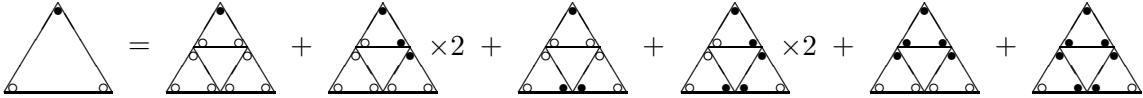
\begin{figure}[htbp]
\unitlength 1.2mm 
\begin{picture}(126,12)
\put(0,0){\line(1,0){12}}
\put(0,0){\line(3,5){6}}
\put(12,0){\line(-3,5){6}}
\multiput(1,0.5)(10,0){2}{\circle{1}}
\put(6,9){\circle*{1}}
\put(15,5){\makebox(0,0){$=$}}
\put(18,0){\line(1,0){12}}
\put(18,0){\line(3,5){6}}
\put(24,0){\line(3,5){3}}
\put(24,0){\line(-3,5){3}}
\put(21,5){\line(1,0){6}}
\put(30,0){\line(-3,5){6}}
\put(19,0.5){\circle{1}}
\put(21,4){\circle{1}}
\put(23,0.5){\circle{1}}
\put(22,5.5){\circle{1}}
\put(24,9){\circle*{1}}
\put(26,5.5){\circle{1}}
\put(25,0.5){\circle{1}}
\put(27,4){\circle{1}}
\put(29,0.5){\circle{1}}
\put(33,5){\makebox(0,0){$+$}}
\put(36,0){\line(1,0){12}}
\put(36,0){\line(3,5){6}}
\put(42,0){\line(3,5){3}}
\put(42,0){\line(-3,5){3}}
\put(39,5){\line(1,0){6}}
\put(48,0){\line(-3,5){6}}
\put(37,0.5){\circle{1}}
\put(39,4){\circle{1}}
\put(41,0.5){\circle{1}}
\put(40,5.5){\circle{1}}
\put(42,9){\circle*{1}}
\put(44,5.5){\circle*{1}}
\put(43,0.5){\circle{1}}
\put(45,4){\circle*{1}}
\put(47,0.5){\circle{1}}
\put(49,5){\makebox(0,0){$\times 2$}}
\put(54,5){\makebox(0,0){$+$}}
\put(57,0){\line(1,0){12}}
\put(57,0){\line(3,5){6}}
\put(63,0){\line(3,5){3}}
\put(63,0){\line(-3,5){3}}
\put(60,5){\line(1,0){6}}
\put(69,0){\line(-3,5){6}}
\put(58,0.5){\circle{1}}
\put(60,4){\circle{1}}
\put(62,0.5){\circle*{1}}
\put(61,5.5){\circle{1}}
\put(63,9){\circle*{1}}
\put(65,5.5){\circle{1}}
\put(64,0.5){\circle*{1}}
\put(66,4){\circle{1}}
\put(68,0.5){\circle{1}}
\put(72,5){\makebox(0,0){$+$}}
\put(75,0){\line(1,0){12}}
\put(75,0){\line(3,5){6}}
\put(81,0){\line(3,5){3}}
\put(81,0){\line(-3,5){3}}
\put(78,5){\line(1,0){6}}
\put(87,0){\line(-3,5){6}}
\put(76,0.5){\circle{1}}
\put(78,4){\circle{1}}
\put(80,0.5){\circle*{1}}
\put(79,5.5){\circle{1}}
\put(81,9){\circle*{1}}
\put(83,5.5){\circle*{1}}
\put(82,0.5){\circle*{1}}
\put(84,4){\circle*{1}}
\put(86,0.5){\circle{1}}
\put(88,5){\makebox(0,0){$\times 2$}}
\put(93,5){\makebox(0,0){$+$}}
\put(96,0){\line(1,0){12}}
\put(96,0){\line(3,5){6}}
\put(102,0){\line(3,5){3}}
\put(102,0){\line(-3,5){3}}
\put(99,5){\line(1,0){6}}
\put(108,0){\line(-3,5){6}}
\put(97,0.5){\circle{1}}
\put(99,4){\circle*{1}}
\put(101,0.5){\circle{1}}
\put(100,5.5){\circle*{1}}
\put(102,9){\circle*{1}}
\put(104,5.5){\circle*{1}}
\put(103,0.5){\circle{1}}
\put(105,4){\circle*{1}}
\put(107,0.5){\circle{1}}
\put(111,5){\makebox(0,0){$+$}}
\put(114,0){\line(1,0){12}}
\put(114,0){\line(3,5){6}}
\put(120,0){\line(3,5){3}}
\put(120,0){\line(-3,5){3}}
\put(117,5){\line(1,0){6}}
\put(126,0){\line(-3,5){6}}
\put(115,0.5){\circle{1}}
\put(117,4){\circle*{1}}
\put(119,0.5){\circle*{1}}
\put(118,5.5){\circle*{1}}
\put(120,9){\circle*{1}}
\put(122,5.5){\circle*{1}}
\put(121,0.5){\circle*{1}}
\put(123,4){\circle*{1}}
\put(125,0.5){\circle{1}}
\end{picture}

\caption{\footnotesize{Illustration for the expression of $g_2(n+1)$. The multiplication of two on the right-hand-side corresponds to the reflection symmetry with respect to the central vertical axis.}} 
\label{gfig}
\end{figure}

\begin{figure}[htbp]
\unitlength 1.2mm 
\begin{picture}(90,12)
\put(0,0){\line(1,0){12}}
\put(0,0){\line(3,5){6}}
\put(12,0){\line(-3,5){6}}
\multiput(1,0.5)(10,0){2}{\circle*{1}}
\put(6,9){\circle*{1}}
\put(15,5){\makebox(0,0){$=$}}
\put(18,0){\line(1,0){12}}
\put(18,0){\line(3,5){6}}
\put(24,0){\line(3,5){3}}
\put(24,0){\line(-3,5){3}}
\put(21,5){\line(1,0){6}}
\put(30,0){\line(-3,5){6}}
\put(19,0.5){\circle*{1}}
\put(21,4){\circle{1}}
\put(23,0.5){\circle{1}}
\put(22,5.5){\circle{1}}
\put(24,9){\circle*{1}}
\put(26,5.5){\circle{1}}
\put(25,0.5){\circle{1}}
\put(27,4){\circle{1}}
\put(29,0.5){\circle*{1}}
\put(33,5){\makebox(0,0){$+$}}
\put(36,0){\line(1,0){12}}
\put(36,0){\line(3,5){6}}
\put(42,0){\line(3,5){3}}
\put(42,0){\line(-3,5){3}}
\put(39,5){\line(1,0){6}}
\put(48,0){\line(-3,5){6}}
\put(37,0.5){\circle*{1}}
\put(39,4){\circle{1}}
\put(41,0.5){\circle*{1}}
\put(40,5.5){\circle{1}}
\put(42,9){\circle*{1}}
\put(44,5.5){\circle{1}}
\put(43,0.5){\circle*{1}}
\put(45,4){\circle{1}}
\put(47,0.5){\circle*{1}}
\put(49,5){\makebox(0,0){$\times 3$}}
\put(54,5){\makebox(0,0){$+$}}
\put(57,0){\line(1,0){12}}
\put(57,0){\line(3,5){6}}
\put(63,0){\line(3,5){3}}
\put(63,0){\line(-3,5){3}}
\put(60,5){\line(1,0){6}}
\put(69,0){\line(-3,5){6}}
\put(58,0.5){\circle*{1}}
\put(60,4){\circle*{1}}
\put(62,0.5){\circle{1}}
\put(61,5.5){\circle*{1}}
\put(63,9){\circle*{1}}
\put(65,5.5){\circle*{1}}
\put(64,0.5){\circle{1}}
\put(66,4){\circle*{1}}
\put(68,0.5){\circle*{1}}
\put(70,5){\makebox(0,0){$\times 3$}}
\put(75,5){\makebox(0,0){$+$}}
\put(78,0){\line(1,0){12}}
\put(78,0){\line(3,5){6}}
\put(84,0){\line(3,5){3}}
\put(84,0){\line(-3,5){3}}
\put(81,5){\line(1,0){6}}
\put(90,0){\line(-3,5){6}}
\put(79,0.5){\circle*{1}}
\put(81,4){\circle*{1}}
\put(83,0.5){\circle*{1}}
\put(82,5.5){\circle*{1}}
\put(84,9){\circle*{1}}
\put(86,5.5){\circle*{1}}
\put(85,0.5){\circle*{1}}
\put(87,4){\circle*{1}}
\put(89,0.5){\circle*{1}}
\end{picture}

\caption{\footnotesize{Illustration for the expression of  $p_2(n+1)$. The multiplication of three on the right-hand-side corresponds to the three possible orientations of $SG_2(n+1)$.}} 
\label{tfig}
\end{figure}
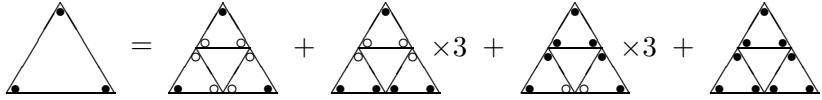

Similarly, $g_2(n+1)$, $h_2(n+1)$ and $p_2(n+1)$ for $SG_2(n+1)$ can be obtained with appropriate configurations of its three constituting $SG_2(n)$ as illustrated in Figs. \ref{gfig}, \ref{hfig} and \ref{tfig} to verify Eqs. (\ref{geq}), (\ref{heq}) and (\ref{peq}), respectively. 
There are always $8=2^3$ terms (counting multiplicity) in Eqs. (\ref{feq}) - (\ref{peq}) because for each of the three pairs of identified vertices it can be either in the vertex subset or not.
\ $\Box$

\bigskip

\begin{figure}[htbp]
\unitlength 1.2mm 
\begin{picture}(126,12)
\put(0,0){\line(1,0){12}}
\put(0,0){\line(3,5){6}}
\put(12,0){\line(-3,5){6}}
\multiput(1,0.5)(10,0){2}{\circle*{1}}
\put(6,9){\circle{1}}
\put(15,5){\makebox(0,0){$=$}}
\put(18,0){\line(1,0){12}}
\put(18,0){\line(3,5){6}}
\put(24,0){\line(3,5){3}}
\put(24,0){\line(-3,5){3}}
\put(21,5){\line(1,0){6}}
\put(30,0){\line(-3,5){6}}
\put(19,0.5){\circle*{1}}
\put(21,4){\circle{1}}
\put(23,0.5){\circle{1}}
\put(22,5.5){\circle{1}}
\put(24,9){\circle{1}}
\put(26,5.5){\circle{1}}
\put(25,0.5){\circle{1}}
\put(27,4){\circle{1}}
\put(29,0.5){\circle*{1}}
\put(33,5){\makebox(0,0){$+$}}
\put(36,0){\line(1,0){12}}
\put(36,0){\line(3,5){6}}
\put(42,0){\line(3,5){3}}
\put(42,0){\line(-3,5){3}}
\put(39,5){\line(1,0){6}}
\put(48,0){\line(-3,5){6}}
\put(37,0.5){\circle*{1}}
\put(39,4){\circle{1}}
\put(41,0.5){\circle*{1}}
\put(40,5.5){\circle{1}}
\put(42,9){\circle{1}}
\put(44,5.5){\circle{1}}
\put(43,0.5){\circle*{1}}
\put(45,4){\circle{1}}
\put(47,0.5){\circle*{1}}
\put(51,5){\makebox(0,0){$+$}}
\put(54,0){\line(1,0){12}}
\put(54,0){\line(3,5){6}}
\put(60,0){\line(3,5){3}}
\put(60,0){\line(-3,5){3}}
\put(57,5){\line(1,0){6}}
\put(66,0){\line(-3,5){6}}
\put(55,0.5){\circle*{1}}
\put(57,4){\circle*{1}}
\put(59,0.5){\circle{1}}
\put(58,5.5){\circle*{1}}
\put(60,9){\circle{1}}
\put(62,5.5){\circle{1}}
\put(61,0.5){\circle{1}}
\put(63,4){\circle{1}}
\put(65,0.5){\circle*{1}}
\put(67,5){\makebox(0,0){$\times 2$}}
\put(72,5){\makebox(0,0){$+$}}
\put(75,0){\line(1,0){12}}
\put(75,0){\line(3,5){6}}
\put(81,0){\line(3,5){3}}
\put(81,0){\line(-3,5){3}}
\put(78,5){\line(1,0){6}}
\put(87,0){\line(-3,5){6}}
\put(76,0.5){\circle*{1}}
\put(78,4){\circle*{1}}
\put(80,0.5){\circle{1}}
\put(79,5.5){\circle*{1}}
\put(81,9){\circle{1}}
\put(83,5.5){\circle*{1}}
\put(82,0.5){\circle{1}}
\put(84,4){\circle*{1}}
\put(86,0.5){\circle*{1}}
\put(90,5){\makebox(0,0){$+$}}
\put(93,0){\line(1,0){12}}
\put(93,0){\line(3,5){6}}
\put(99,0){\line(3,5){3}}
\put(99,0){\line(-3,5){3}}
\put(96,5){\line(1,0){6}}
\put(105,0){\line(-3,5){6}}
\put(94,0.5){\circle*{1}}
\put(96,4){\circle*{1}}
\put(98,0.5){\circle*{1}}
\put(97,5.5){\circle*{1}}
\put(99,9){\circle{1}}
\put(101,5.5){\circle{1}}
\put(100,0.5){\circle*{1}}
\put(102,4){\circle{1}}
\put(104,0.5){\circle*{1}}
\put(106,5){\makebox(0,0){$\times 2$}}
\put(111,5){\makebox(0,0){$+$}}
\put(114,0){\line(1,0){12}}
\put(114,0){\line(3,5){6}}
\put(120,0){\line(3,5){3}}
\put(120,0){\line(-3,5){3}}
\put(117,5){\line(1,0){6}}
\put(126,0){\line(-3,5){6}}
\put(115,0.5){\circle*{1}}
\put(117,4){\circle*{1}}
\put(119,0.5){\circle*{1}}
\put(118,5.5){\circle*{1}}
\put(120,9){\circle{1}}
\put(122,5.5){\circle*{1}}
\put(121,0.5){\circle*{1}}
\put(123,4){\circle*{1}}
\put(125,0.5){\circle*{1}}
\end{picture}

\caption{\footnotesize{Illustration for the expression of $h_2(n+1)$. The multiplication of two on the right-hand-side corresponds to the reflection symmetry with respect to the central vertical axis.}} 
\label{hfig}
\end{figure}
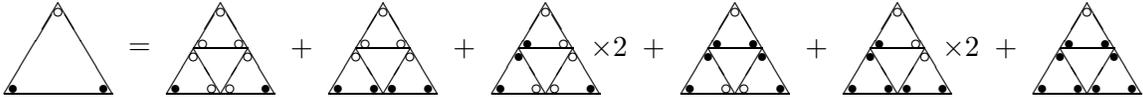

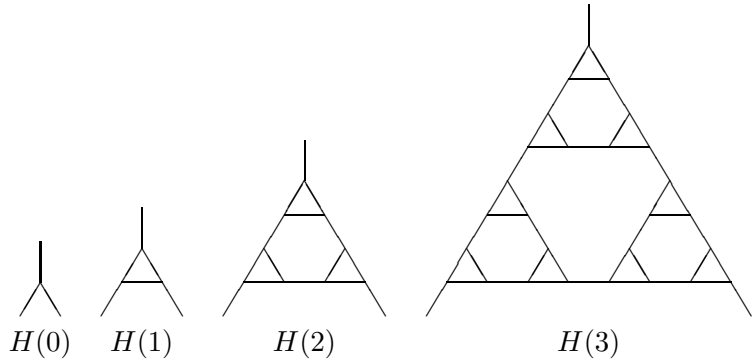
\begin{figure}[htbp]
\unitlength 0.9mm \hspace*{3mm}
\begin{picture}(108,46)
\put(0,0){\line(3,5){3}}
\put(6,0){\line(-3,5){3}}
\put(3,5){\line(0,1){6}}
\put(3,-4){\makebox(0,0){$H(0)$}}
\put(12,0){\line(3,5){6}}
\put(24,0){\line(-3,5){6}}
\put(15,5){\line(1,0){6}}
\put(18,10){\line(0,1){6}}
\put(18,-4){\makebox(0,0){$H(1)$}}
\put(30,0){\line(3,5){12}}
\put(54,0){\line(-3,5){12}}
\put(33,5){\line(1,0){18}}
\put(39,5){\line(-3,5){3}}
\put(45,5){\line(3,5){3}}
\put(39,15){\line(1,0){6}}
\put(42,20){\line(0,1){6}}
\put(42,-4){\makebox(0,0){$H(2)$}}
\put(63,5){\line(1,0){42}}
\put(75,25){\line(1,0){18}}
\put(60,0){\line(3,5){24}}
\put(87,5){\line(3,5){9}}
\put(81,5){\line(-3,5){9}}
\put(108,0){\line(-3,5){24}}
\multiput(69,15)(24,0){2}{\line(1,0){6}}
\multiput(69,5)(24,0){2}{\line(-3,5){3}}
\multiput(75,5)(24,0){2}{\line(3,5){3}}
\put(81,35){\line(1,0){6}}
\put(81,25){\line(-3,5){3}}
\put(87,25){\line(3,5){3}}
\put(84,40){\line(0,1){6}}
\put(84,-4){\makebox(0,0){$H(3)$}}
\end{picture}

\vspace*{5mm}
\caption{\footnotesize{The first four stages $n=0,1,2,3$ of the graph $H(n)$.}} 
\label{sglfig}
\end{figure}

Alternatively, it is known that the number of dimer-monomers on a graph $G$ is the same as the number of independent sets on the associated line graph $L(G)$ \cite{yan}. Consider the sequence of graph $H(n)$ shown in Fig. \ref{sglfig} that is obtained by adding an extra edge to each of the outmost vertices of the Hanoi graph. As $H(n)$ has $SG_2(n)$ as its line graph, the enumeration of the number of independent sets on $SG_2(n)$ is equivalent to the enumeration of the number of dimer-monomers on these $H(n)$. One can define corresponding quantities of $f_2(n)$, $g_2(n)$, $h_2(n)$, $p_2(n)$ on $H(n)$ that satisfy the same recursion relations as in Lemma \ref{lemmasg2r}.

The values of $f_2(n)$, $g_2(n)$, $h_2(n)$, $p_2(n)$ for small $n$ can be evaluated recursively by Eqs. (\ref{feq}) - (\ref{peq}) as listed in Table \ref{tablesg2}. These numbers grow exponentially, and do not have simple integer factorizations. To estimate the value of the asymptotic growth constant defined in Eq. (\ref{zdef}), we need the following lemmas. 

For the generalized two-dimensional Sierpinski gasket $SG_{2,b}(n)$, define the ratios
\beq
\alpha_{2,b}(n) = \frac{g_{2,b}(n)}{f_{2,b}(n)} \ , \qquad \beta_{2,b}(n) = \frac{h_{2,b}(n)}{g_{2,b}(n)} \ , \qquad \gamma_{2,b}(n) = \frac{p_{2,b}(n)}{h_{2,b}(n)} \ ,
\label{ratiodef}
\eeq
where $n$ is a positive integer. For the ordinary Sierpinski gasket in this section, they are simplified to be $\alpha_2(n)$, $\beta_2(n)$, $\gamma_2(n)$ and their values for small $n$ are listed in Table \ref{tablesg2r}. From the initial values of $f_2(n)$, $g_2(n)$, $h_2(n)$, $p_2(n)$, it is easy to see that $f_2(n) \ge g_2(n) \ge h_2(n) \ge p_2(n)$ for all non-negative $n$ by induction. It follows that $\alpha_2(n),\beta_2(n),\gamma_2(n) \in (0,1]$.

\begin{table}[htbp]
\caption{\label{tablesg2} The first few values of $f_2(n)$, $g_2(n)$, $h_2(n)$, $p_2(n)$, $m_2(n)$.}
\begin{center}
\begin{tabular}{|c||r|r|r|r|r|}
\hline\hline 
$n$      & 0 &  1 &   2 &          3 &                           4 \\ \hline\hline 
$f_2(n)$ & 1 &  4 & 125 &  4,007,754 & 132,460,031,222,098,852,477 \\ \hline 
$g_2(n)$ & 1 &  2 &  65 &  2,089,888 &  69,073,020,285,472,159,669 \\ \hline 
$h_2(n)$ & 0 &  1 &  34 &  1,089,805 &  36,019,032,212,213,865,476 \\ \hline 
$p_2(n)$ & 0 &  1 &  18 &    568,301 &  18,782,596,680,434,060,148 \\ \hline
$m_2(n)$ & 4 & 14 & 440 & 14,115,134 & 466,518,785,395,590,988,060 \\ \hline\hline 
\end{tabular}
\end{center}
\end{table}

\begin{table}[htbp]
\caption{\label{tablesg2r} The first few values of $\alpha_2(n)$, $\beta_2(n)$, $\gamma_2(n)$. The last digits given are rounded off.}
\begin{center}
\begin{tabular}{|c||r|r|r|r|r|}
\hline\hline 
$n$           &   1 &                 2 &                 3 &                 4 \\ \hline\hline 
$\alpha_2(n)$ & 0.5 & 0.52              & 0.521461147565444 & 0.521463113425180 \\ \hline 
$\beta_2(n)$  & 0.5 & 0.523076923076923 & 0.521465743618797 & 0.521463113431998 \\ \hline 
$\gamma_2(n)$ &   1 & 0.529411764705882 & 0.521470354788242 & 0.521463113438816 \\ \hline\hline
\end{tabular}
\end{center}
\end{table}

\bigskip

\begin{lem} \label{lemmasg2abc} For any positive integer $n$, the ratios satisfy
\beq
\alpha_2(n) \le \beta_2(n) \le \gamma_2(n) \ .
\label{abc}
\eeq
When $n$ increases, the ratio $\alpha_2(n)$ increases monotonically while $\gamma_2(n)$ decreases monotonically. The three ratios in the large $n$ limit are equal to each other
\beq
\lim _{n \to \infty} \alpha_2(n) = \lim _{n \to \infty} \beta_2(n) = \lim _{n \to \infty} \gamma_2(n) \ .
\eeq
\end{lem}

{\sl Proof} \quad 
It is clear that Eq. (\ref{abc}) is valid for small values of $n$ given in Table {\ref{tablesg2r}}. In order to save space, we will use $\alpha_n$, $\beta_n$, $\gamma_n$ to denote $\alpha_2(n)$, $\beta_2(n)$, $\gamma_2(n)$ for the lengthy equations in this Lemma. By definition, we have
\beq
\alpha_{n+1} = \alpha_n \frac{B_n}{A_n} \ , \qquad 
\beta_{n+1} = \alpha_n \frac{C_n}{B_n} \ , \qquad 
\gamma_{n+1} = \alpha_n \frac{D_n}{C_n} 
\eeq
for a positive $n$, where 
\beqs
A_n & = & 1 + 3\alpha_n^2 + 3\alpha_n^3\beta_n + \alpha_n^3\beta_n^3 \ , \cr
B_n & = & 1 + 2\alpha_n\beta_n + \alpha_n^2 + 2\alpha_n^2\beta_n^2 + \alpha_n^2\beta_n\gamma_n + \alpha_n^2\beta_n^3\gamma_n \ , \cr
C_n & = & 1 + \beta_n^2 + 2\alpha_n\beta_n + \alpha_n\beta_n^3 + 2\alpha_n\beta_n^2\gamma_n + \alpha_n\beta_n^3\gamma_n^2 \ , \cr
D_n & = & 1 + 3\beta_n^2 + 3\beta_n^3\gamma_n + \beta_n^3\gamma_n^3 \ ,
\eeqs
such that
\beq
\alpha_{n+1} - \alpha_n = \frac{1}{A_n} \Bigl\{ 2\alpha_n^2 (1+\alpha_n\beta_n) (\beta_n-\alpha_n) + \alpha_n^3\beta_n (1+\beta_n^2) (\gamma_n-\alpha_n) \Bigr\} \ , 
\label{alphad}
\eeq
\beq
\beta_{n+1}-\alpha_n = \frac{1}{B_n} \Bigl\{ \alpha_n (\beta_n+\alpha_n+\alpha_n\beta_n^2+\alpha_n\beta_n\gamma_n) (\beta_n-\alpha_n) + \alpha_n^2\beta_n^2 (1+\beta_n\gamma_n) (\gamma_n-\alpha_n) \Bigr\} \ .
\eeq
It follows that 
\beqs
\lefteqn{\beta_{n+1}-\alpha_{n+1}} \cr & = & \frac{\beta_n-\alpha_n}{A_nB_n} \Bigl\{ \alpha_n (\beta_n+\alpha_n+\alpha_n\beta_n^2+\alpha_n\beta_n\gamma_n) A_n - 2\alpha_n^2 (1+\alpha_n\beta_n) B_n \Bigr\} \cr
& & + \frac{\gamma_n-\alpha_n}{A_nB_n} \Bigl\{ \alpha_n^2\beta_n^2 (1+\beta_n\gamma_n) A_n - \alpha_n^3\beta_n (1+\beta_n^2) B_n \Bigr\} \cr 
& = & \frac{\beta_n-\alpha_n}{A_nB_n} \alpha_n (1+\alpha_n\beta_n) \Bigl\{ (1-\alpha_n^2-\alpha_n^2\beta_n\gamma_n) (\beta_n-\alpha_n) + \alpha_n\beta_n (\gamma_n-\alpha_n) - \alpha_n^3\beta_n^3 (\gamma_n-\beta_n) \Bigr\} \cr
& & + \frac{\gamma_n-\alpha_n}{A_nB_n} \alpha_n^2\beta_n \Bigl\{ (1+\alpha_n^2+\alpha_n^2\beta_n\gamma_n) (\beta_n-\alpha_n) + \beta_n^2 (\gamma_n-\alpha_n) + \alpha_n^2\beta_n^2 (2+\alpha_n\beta_n) (\gamma_n-\beta_n) \Bigr\} \ , \cr
& &  
\eeqs
where 
\beqs
A_nB_n & = & 1 + 4\alpha_n^2 + 2\alpha_n\beta_n + 3\alpha_n^4 + 9\alpha_n^3\beta_n + 2\alpha_n^2\beta_n^2 + \alpha_n^2\beta_n\gamma_n + 3\alpha_n^5\beta_n + 12\alpha_n^4\beta_n^2 + \alpha_n^3\beta_n^3 \cr
& & + 3\alpha_n^4\beta_n\gamma_n + \alpha_n^2\beta_n^3\gamma_n + 7\alpha_n^5\beta_n^3 + 2\alpha_n^4\beta_n^4 + 3\alpha_n^5\beta_n^2\gamma_n + 3\alpha_n^4\beta_n^3\gamma_n + 2\alpha_n^5\beta_n^5 + 4\alpha_n^5\beta_n^4\gamma_n \cr
& & + \alpha_n^5\beta_n^6\gamma_n \ .
\eeqs
Using the fact that $\alpha_n,\beta_n,\gamma_n \in (0,1]$ and the inequality $\beta_n \le \gamma_n$ to be shown below, $\alpha_n \le \beta_n$ is proved by induction. Define $\epsilon_n=\gamma_n-\alpha_n$, which is larger than $\gamma_n-\beta_n$ and $\beta_n-\alpha_n$ as we shall prove $\beta_n \le \gamma_n$, then
\beqs
\beta_{n+1}-\alpha_{n+1} & \le & \frac{\epsilon_n^2}{A_nB_n} \Bigl\{ \alpha_n (1 + \alpha_n\beta_n)^2 + \alpha_n^2\beta_n (1 + \alpha_n^2 + \alpha_n^2\beta_n\gamma_n + \beta_n^2 + 2\alpha_n^2\beta_n^2 + \alpha_n^3\beta_n^3) \Bigr\} \cr
& = & \frac{\epsilon_n^2}{A_nB_n} \Bigl\{ \alpha_n + 3\alpha_n^2\beta_n + \alpha_n^3\beta_n^2 + \alpha_n^4\beta_n + \alpha_n^2\beta_n^3 + 2\alpha_n^4\beta_n^3 + \alpha_n^4\beta_n^2\gamma_n + \alpha_n^5\beta_n^4 \Bigr\} \cr
& \le & \frac{\epsilon_n^2}{A_nB_n} \Bigl\{ 1 + 3\alpha_n^2 + 2\alpha_n^3\beta_n + \alpha_n^2\beta_n^2 + 3\alpha_n^4\beta_n^2 + \alpha_n^5\beta_n^3 \Bigr\} \le \epsilon_n^2 \ .
\label{betamalpha}
\eeqs

Similarly, we have
\beq
\gamma_{n+1}-\alpha_n = \frac{1}{C_n} \Bigl\{ 2\alpha_n\beta_n (1+\beta_n\gamma_n) (\beta_n-\alpha_n) + \alpha_n\beta_n^3 (1+\gamma_n^2) (\gamma_n-\alpha_n) \Bigr\} \ ,
\eeq
and
\beqs
\gamma_{n+1}-\beta_{n+1} & = & \frac{\beta_n-\alpha_n}{B_nC_n} \Bigl\{ 2\alpha_n\beta_n (1+\beta_n\gamma_n) B_n - \alpha_n (\beta_n+\alpha_n+\alpha_n\beta_n^2+\alpha_n\beta_n\gamma_n) C_n \Bigr\} \cr
& & + \frac{\gamma_n-\alpha_n}{B_nC_n} \Bigl\{ \alpha_n\beta_n^3 (1+\gamma_n^2) B_n - \alpha_n^2\beta_n^2 (1+\beta_n\gamma_n) C_n \Bigr\} \cr 
& = & \frac{\beta_n-\alpha_n}{B_nC_n} \alpha_n \Bigl\{ (1+\beta_n\gamma_n-\alpha_n\beta_n^3) (\beta_n-\alpha_n) \cr
& & + \beta_n (\beta_n+\alpha_n\beta_n^2+\alpha_n^2\beta_n^3-\alpha_n^2\beta_n^2\gamma_n-\alpha_n^2\beta_n^3\gamma_n^2) (\gamma_n-\beta_n) - \alpha_n\beta_n^4\gamma_n (\gamma_n-\alpha_n) \Bigr\} \cr
& & + \frac{\gamma_n-\alpha_n}{B_nC_n} \alpha_n\beta_n^2 \Bigl\{ (1+\alpha_n\beta_n) (\beta_n-\alpha_n) + \beta_n (\gamma_n-\alpha_n\beta_n^2+2\alpha_n\beta_n\gamma_n) (\gamma_n-\alpha_n) \cr
& & + \alpha_n^2\beta_n\gamma_n (1+\beta_n\gamma_n) (\gamma_n-\beta_n) \Bigr\} \ ,
\eeqs
where 
\beqs
B_nC_n & = & 1 + \alpha_n^2 + 4\alpha_n\beta_n + \beta_n^2 + 2\alpha_n^3\beta_n + 7\alpha_n^2\beta_n^2 + 3\alpha_n\beta_n^3 + \alpha_n^2\beta_n\gamma_n + 2\alpha_n\beta_n^2\gamma_n + 5\alpha_n^3\beta_n^3 \cr
& & + 4\alpha_n^2\beta_n^4 + 4\alpha_n^3\beta_n^2\gamma_n + 6\alpha_n^2\beta_n^3\gamma_n + \alpha_n\beta_n^3\gamma_n^2 + 2\alpha_n^3\beta_n^5 + 7\alpha_n^3\beta_n^4\gamma_n + \alpha_n^2\beta_n^5\gamma_n \cr
& & + 3\alpha_n^3\beta_n^3\gamma_n^2 + 2\alpha_n^2\beta_n^4\gamma_n^2 + \alpha_n^3\beta_n^6\gamma_n + 4\alpha_n^3\beta_n^5\gamma_n^2 + \alpha_n^3\beta_n^4\gamma_n^3 + \alpha_n^3\beta_n^6\gamma_n^3 \ .
\eeqs
Using the fact that $\alpha_n,\beta_n,\gamma_n \in (0,1]$ and the inequality $\alpha_n \le \beta_n$ given above, $\beta_n \le \gamma_n$ is proved by induction. We also have
\beqs
\gamma_{n+1}-\beta_{n+1} & \le & \frac{\epsilon_n^2}{B_nC_n} \Bigl\{ \alpha_n (1 + \beta_n\gamma_n + \beta_n^2 + \alpha_n\beta_n^3 + \alpha_n^2\beta_n^4) \cr
& & + \alpha_n\beta_n^2 (1 + \alpha_n\beta_n + \beta_n\gamma_n + 2\alpha_n\beta_n^2\gamma_n + \alpha_n^2\beta_n\gamma_n + \alpha_n^2\beta_n^2\gamma_n^2) \Bigr\} \cr
& = & \frac{\epsilon_n^2}{B_nC_n} \Bigl\{ \alpha_n + 2\alpha_n\beta_n^2 + \alpha_n\beta_n\gamma_n + 2\alpha_n^2\beta_n^3 + \alpha_n^3\beta_n^4 + \alpha_n\beta_n^3\gamma_n + \alpha_n^3\beta_n^3\gamma_n \cr
& & + 2\alpha_n^2\beta_n^4\gamma_n + \alpha_n^3\beta_n^4\gamma_n^2 \Bigr\} \cr
& \le & \frac{\epsilon_n^2}{B_nC_n} \Bigl\{ 1 + 3\alpha_n\beta_n + 3\alpha_n\beta_n^3 + 2\alpha_n^3\beta_n^3 + 2\alpha_n^2\beta_n^3\gamma_n + \alpha_n^3\beta_n^4\gamma_n \Bigr\} \le \epsilon_n^2 \ .
\label{gammambeta}
\eeqs
From Eqs. (\ref{betamalpha}) and (\ref{gammambeta}), we obtain $\epsilon_{n+1} \leq 2\epsilon_n^2$ for all positive $n$ by induction. It follows that for any positive integer $m \le n$,
\beq
\epsilon_n \leq 2\epsilon_{n-1}^2 \leq 2 \big[ 2\epsilon_{n-2}^2 \big]^2 \leq \cdots \leq \frac12 \big[ 2\epsilon_m \big]^{2^{n-m}}.
\eeq
Taking $m$ as an integer larger than one, then the values of $\alpha_n$, $\beta_n$, $\gamma_n$ are close to each other when $n$ becomes large.

Finally, it is clear that $\alpha_2(n)$ increases monotonically as $n$ increases by Eq. (\ref{alphad}). As 
\beq
\gamma_n - \gamma_{n+1} = \frac{1}{C_n} \Bigl\{ (1+\beta_n^2) (\gamma_n-\alpha_n) + 2\alpha_n\beta_n (1+\beta_n\gamma_n) (\gamma_n-\beta_n) \Bigr\} \ , 
\eeq
we know $\gamma_2(n)$ decreases monotonically as $n$ increases, and the proof is completed. 
\ $\Box$

\bigskip

Numerically, we find
\beq
\lim _{n \to \infty} \alpha_2(n) = \lim _{n \to \infty} \beta_2(n) = \lim _{n \to \infty} \gamma_2(n) = 0.521463113428094965776...
\eeq
From the above lemma, we have the following bounds for the asymptotic growth constant.

\bigskip

\begin{lem} \label{lemmasg2b} The asymptotic growth constant for the number of independent sets on $SG_2(n)$ is bounded:
\beq
\frac{2}{3^{m+1}} \ln [f_2(m)] + \frac{1}{3^m} \ln [1+\alpha_2^2(m)] \le z_{SG_2} \le \frac{2}{3^{m+1}} \ln [f_2(m)] + \frac{1}{3^m} \ln [1+\gamma_2^2(m)] \ ,
\label{zsg2}
\eeq
where $m$ is a positive integer.
\end{lem}

{\sl Proof} \quad 
From Eq. (\ref{feq}) and Lemma \ref{lemmasg2abc}, we have the upper bound for $f_2(n)$,
\beqs
f_2(n) & = & f_2^3(n-1) \big[ 1 + 3\alpha_2^2(n-1) + 3\alpha_2^3(n-1)\beta_2(n-1) + \alpha_2^3(n-1)\beta_2^3(n-1) \big] \cr 
& \le & f_2^3(n-1) \big[ 1 + \gamma_2^2(n-1) \big]^3 \cr
& \le & \Bigl \{ f_2^3(n-2) \big[ 1 + \gamma_2^2(n-2) \big]^3 \Bigr \}^3 \big[ 1 + \gamma_2^2(n-1) \big]^3 \le \cdots \cr
& \le & \big[ f_2(m) \big]^{3^{n-m}} \big[ 1 + \gamma_2^2(m) \big]^{\frac32 (3^{n-m}-1)} \ .
\eeqs
From Eq. (\ref{meq}), the number of independent sets has the upper bound
\beqs
m_2(n) & = & f_2(n) \big[ 1 + 3\alpha_2(n) + 3\alpha_2(n)\beta_2(n) + \alpha_2(n)\beta_2(n)\gamma_2(n) \big] \cr
& \le & \big[ f_2(m) \big]^{3^{n-m}} \big[ 1 + \gamma_2^2(m) \big]^{\frac32 (3^{n-m}-1)} \big[ 1 + \gamma_2(n) \big]^3 \ .
\eeqs
As the number of vertices of $SG_2(n)$ is $3(3^n+1)/2$ by Eq. (\ref{v}), the upper bound for $z_{SG_2}$ defined in Eq. (\ref{zdef}) follows. The lower bound for $z_{SG_2}$ can be derived similarly.
\ $\Box$

\bigskip

As $m$ increases, the difference between the upper and lower bounds in Eq. (\ref{zsg2}) becomes small and the convergence is rapid. The numerical value of $z_{SG_2}$ can be obtained with more than a hundred significant figures accurate when $m$ is equal to eight. 

\bigskip

\begin{propo} \label{proposg2} The asymptotic growth constant for the number of independent sets on the two-dimensional Sierpinski gasket $SG_2(n)$ in the large $n$ limit is $z_{SG_2}=0.38430953443368558352...$.

\end{propo}

\bigskip

As mentioned previously, the number of dimer-monomers on the graph $H(n)$ illustrated in Fig. \ref{sglfig} is the same as the number of independent sets on the two-dimensional Sierpinski gasket $SG_2(n)$. Similar to Eq. (\ref{zdef}), one can define a constant for the exponential growth of the number of dimer-monomers:
\beq
z^\prime_G = \lim_{v(G) \to \infty} \frac{\ln N_{DM}(G)}{v(G)} \ ,
\label{zdefn}
\eeq
where $N_{DM}(G)$ is the number of dimer-monomers on a graph $G$. As the number of vertices of $H(n)$ is $3^n+3$, we have the following corollary.

\bigskip

\begin{cor} \label{corsg2} The asymptotic growth constant for the number of dimer-monomers on the graph $H(n)$ in the large $n$ limit is $z^\prime_{H}=0.57646430165052837528...$.

\end{cor}

\bigskip

To the best of our knowledge, this result has not been obtained elsewhere.

\section{The number of independent sets on $SG_{2,3}(n)$} 

The method given in the previous section can be applied to the number of independent sets on $SG_{d,b}(n)$ with larger values of $d$ and $b$. The number of configurations to be considered increases as $d$ and $b$ increase, and the recursion relations must be derived individually for each $d$ and $b$. 
In this section, we consider the generalized two-dimensional Sierpinski gasket $SG_{2,b}(n)$ with the number of layers $b$ equal to three. 
For $SG_{2,3}(n)$, the numbers of edges and vertices are given by 
\beq
e(SG_{2,3}(n)) = 3 \times 6^n \ ,
\label{esg23}
\eeq
\beq
v(SG_{2,3}(n)) = \frac{7 \times 6^n + 8}{5} \ ,
\label{vsg23}
\eeq
where the three outmost vertices have degree two. There are $(6^n-1)/5$ vertices of $SG_{2,3}(n)$ with degree six and $6(6^n-1)/5$ vertices with degree four. The initial values for the number of independent sets with various conditions are the same as those for $SG_2$: $f_{2,3}(0)=1$, $g_{2,3}(0)=1$, $h_{2,3}(0)=0$ and $p_{2,3}(0)=0$. The recursion relations for $SG_{2,3}(n)$ are lengthy and given in the appendix. Some values of $f_{2,3}(n)$, $g_{2,3}(n)$, $h_{2,3}(n)$, $p_{2,3}(n)$, $m_{2,3}(n)$ are listed in Table \ref{tablesg23}. These numbers grow exponentially, and do not have simple integer factorizations.

\begin{table}[htbp]
\caption{\label{tablesg23} The first few values of $f_{2,3}(n)$, $g_{2,3}(n)$, $h_{2,3}(n)$, $p_{2,3}(n)$, $m_{2,3}(n)$.}
\begin{center}
\begin{tabular}{|c||r|r|r|r|}
\hline\hline 
$n$          & 0 &  1 &           2 & 3 \\ \hline\hline 
$f_{2,3}(n)$ & 1 & 19 & 172,371,175 & 93,818,345,014,803,648,739,612,995,034,820,933,103,277,876,214,071 \\ \hline 
$g_{2,3}(n)$ & 1 &  9 &  80,291,169 & 43,700,938,182,461,202,772,695,141,988,444,331,720,442,482,282,619 \\ \hline 
$h_{2,3}(n)$ & 0 &  4 &  37,399,906 & 20,356,061,468,851,869,739,344,457,713,631,919,274,541,443,648,604 \\ \hline 
$p_{2,3}(n)$ & 0 &  2 &  17,420,990 & 9,481,930,039,890,479,716,613,035,420,873,292,623,048,215,623,126 \\ \hline
$m_{2,3}(n)$ & 4 & 60 & 542,865,390 & 295,471,274,008,633,345,992,344,829,561,922,978,711,277,869,630,866 \\ \hline\hline 
\end{tabular}
\end{center}
\end{table}

The values of the ratios $\alpha_{2,3}(n)$, $\beta_{2,3}(n)$, $\gamma_{2,3}(n)$ defined in Eq. (\ref{ratiodef}) for small $n$ are listed in Table \ref{tablesg23n}. 
The sequence of $\alpha_{2,3}(n)$ decreases monotonically as $n$ increases, while $\beta_{2,3}(n)$ increases monotonically. Except the first term $\gamma_{2,3}(1)$, $\gamma_{2,3}(n)$ also increases monotonically for $n \ge 2$. We again have $\alpha_{2,3}(n), \beta_{2,3}(n), \gamma_{2,3}(n) \in (0,1]$ but $\gamma_{2,3}(n) \le \beta_{2,3}(n) \le \alpha_{2,3}(n)$ for $n \ge 2$, in contrast to Lemma \ref{lemmasg2abc}.

\begin{table}[htbp]
\caption{\label{tablesg23n} The first few values of $\alpha_{2,3}(n)$, $\beta_{2,3}(n)$, $\gamma_{2,3}(n)$. The last digits given are rounded off.}
\begin{center}
\begin{tabular}{|c||r|r|r|}
\hline\hline 
$n$               &                      1 &                      2 & 3  \\ \hline\hline 
$\alpha_{2,3}(n)$ & 0.47368421052631578947 & 0.46580391994195085112 & 0.46580376338514186621 \\ \hline 
$\beta_{2,3}(n)$  & 0.44444444444444444444 & 0.46580348082863259844 & 0.46580376338514186620 \\ \hline
$\gamma_{2,3}(n)$ & 0.5                    & 0.46580304239267339335 & 0.46580376338514186619 \\ \hline\hline 
\end{tabular}
\end{center}
\end{table}

By the same argument given in Lemma \ref{lemmasg2b}, we have the upper and lower bounds of the asymptotic growth constant for the number of independent sets on $SG_{2,3}(n)$:
\beqs
& & \frac{1}{7\times 6^m} \Bigl\{ 5\ln f_{2,3}(m) + \ln \big[ 1+\gamma_{2,3}^3(m) \big] + 6\ln \big[1+\gamma_{2,3}^2(m) \big] \Bigr\} \le z_{SG_{2,3}} \cr\cr
& & \le \frac{1}{7\times 6^m} \Bigl\{ 5\ln f_{2,3}(m) + \ln \big[ 1+\alpha_{2,3}^3(m) \big] + 6\ln \big[1+\alpha_{2,3}^2(m) \big] \Bigr\}  \ ,
\label{zsg23}
\eeqs
with $m$ a positive integer. The convergence of the upper and lower bounds remains rapid. More than a hundred significant figures for $z_{SG_{2,3}}$ can be obtained when $m$ is equal to five. We have the following proposition.

\bigskip

\begin{propo} \label{proposg23} The asymptotic growth constant for the number of independent sets on the two-dimensional Sierpinski gasket $SG_{2,3}(n)$ in the large $n$ limit is $z_{SG_{2,3}}=0.38135033366164857274...$.

\end{propo}

\section{The number of independent sets on $SG_3(n)$} 

In this section, we derive the asymptotic growth constant of independent sets on the three-dimensional Sierpinski gasket $SG_3(n)$. We use the following definitions.

\bigskip

\begin{defi} \label{defisg3} Consider the three-dimensional Sierpinski gasket $SG_3(n)$ at stage $n$. (i) Define $m_3(n) \equiv N_{IS}(SG_3(n))$ as the number of independent sets. (ii) Define $f_3(n)$ as the number of independent sets such that all four outmost vertices are not in the vertex subset. (iii) Define $g_3(n)$ as the number of independent sets such that one certain outmost vertex is in the vertex subset. (iv) Define $h_3(n)$ as the number of independent sets such that two certain outmost vertices are in the vertex subset. (v) Define $p_3(n)$ as the number of independent sets such that three certain outmost vertices are in the vertex subset. (vi) Define $q_3(n)$ as the number of independent sets such that all four outmost vertices are in the vertex subset.
\end{defi}

\bigskip

\begin{figure}[htbp]
\unitlength 1.8mm 
\begin{picture}(54,5)
\put(0,0){\line(1,0){6}}
\put(0,0){\line(3,5){3}}
\put(6,0){\line(-3,5){3}}
\put(0,0){\line(3,2){3}}
\put(6,0){\line(-3,2){3}}
\put(3,2){\line(0,1){3}}
\put(1,0.5){\circle{1}}
\put(5,0.5){\circle{1}}
\put(3,2){\circle{1}}
\put(3,4){\circle{1}}
\put(3,-2){\makebox(0,0){$f_3(n)$}}
\put(12,0){\line(1,0){6}}
\put(12,0){\line(3,5){3}}
\put(18,0){\line(-3,5){3}}
\put(12,0){\line(3,2){3}}
\put(18,0){\line(-3,2){3}}
\put(15,2){\line(0,1){3}}
\put(13,0.5){\circle{1}}
\put(17,0.5){\circle{1}}
\put(15,2){\circle*{1}}
\put(15,4){\circle{1}}
\put(15,-2){\makebox(0,0){$g_3(n)$}}
\put(24,0){\line(1,0){6}}
\put(24,0){\line(3,5){3}}
\put(30,0){\line(-3,5){3}}
\put(24,0){\line(3,2){3}}
\put(30,0){\line(-3,2){3}}
\put(27,2){\line(0,1){3}}
\put(25,0.5){\circle*{1}}
\put(29,0.5){\circle*{1}}
\put(27,2){\circle{1}}
\put(27,4){\circle{1}}
\put(27,-2){\makebox(0,0){$h_3(n)$}}
\put(36,0){\line(1,0){6}}
\put(36,0){\line(3,5){3}}
\put(42,0){\line(-3,5){3}}
\put(36,0){\line(3,2){3}}
\put(42,0){\line(-3,2){3}}
\put(39,2){\line(0,1){3}}
\put(37,0.5){\circle*{1}}
\put(41,0.5){\circle*{1}}
\put(39,2){\circle{1}}
\put(39,4){\circle*{1}}
\put(39,-2){\makebox(0,0){$p_3(n)$}}
\put(48,0){\line(1,0){6}}
\put(48,0){\line(3,5){3}}
\put(54,0){\line(-3,5){3}}
\put(48,0){\line(3,2){3}}
\put(54,0){\line(-3,2){3}}
\put(51,2){\line(0,1){3}}
\put(49,0.5){\circle*{1}}
\put(53,0.5){\circle*{1}}
\put(51,2){\circle*{1}}
\put(51,4){\circle*{1}}
\put(51,-2){\makebox(0,0){$q_3(n)$}}
\end{picture}

\vspace*{5mm}
\caption{\footnotesize{Illustration for the configurations $f_3(n)$, $g_3(n)$, $h_3(n)$, $p_3(n)$ and $q_3(n)$. Only the four outmost vertices are shown explicitly, where a solid circle is in the vertex subset and an open circle is not.}} 
\label{fghpqfig}
\end{figure}
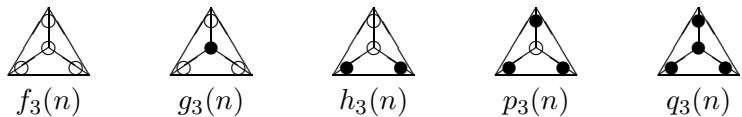

The quantities $f_3(n)$, $g_3(n)$, $h_3(n)$, $p_3(n)$ and $q_3(n)$ are illustrated in Fig. \ref{fghpqfig}, where only the outmost vertices are shown. There are ${4 \choose 1}=4$ equivalent $g_3(n)$, ${4 \choose 2}=6$ equivalent $h_3(n)$, and ${4 \choose 1}=4$ equivalent $p_3(n)$. By definition,
\beq
m_3(n) = f_3(n)+4g_3(n)+6h_3(n)+4p_3(n)+q_3(n) \ .
\label{fsg3}
\eeq
The initial values at stage zero are $f_3(0)=1$, $g_3(0)=1$, $h_3(0)=0$, $p_3(0)=0$, $q_3(0)=0$ and $m_3(0)=5$.

The recursion relations are lengthy and given in the appendix. Some values of $f_3(n)$, $g_3(n)$, $h_3(n)$, $p_3(n)$, $q_3(n)$, $m_3(n)$ are listed in Table \ref{tablesg3}. These numbers grow exponentially, and do not have simple integer factorizations.

\begin{table}[htbp]
\caption{\label{tablesg3} The first few values of $f_3(n)$, $g_3(n)$, $h_3(n)$, $p_3(n)$, $q_3(n)$, $m_3(n)$.}
\begin{center}
\begin{tabular}{|c||r|r|r|r|}
\hline\hline 
$n$      & 0 &  1 &       2 & 3 \\ \hline\hline 
$f_3(n)$ & 1 & 10 &  25,817 & 1,292,964,293,737,151,090 \\ \hline 
$g_3(n)$ & 1 &  4 &  11,387 &   571,820,791,550,665,532 \\ \hline 
$h_3(n)$ & 0 &  2 &   5,050 &   252,892,039,471,313,074 \\ \hline
$p_3(n)$ & 0 &  1 &   2,252 &   111,843,868,747,687,217 \\ \hline 
$q_3(n)$ & 0 &  1 &   1,010 &    49,464,202,269,253,193 \\ \hline
$m_3(n)$ & 5 & 43 & 111,683 & 5,594,439,374,027,693,723 \\ \hline\hline 
\end{tabular}
\end{center}
\end{table}

Define ratios
\beq
\alpha_3(n)=\frac{g_3(n)}{f_3(n)} \ , \qquad \beta_3(n)=\frac{h_3(n)}{g_3(n)} \ , \qquad \gamma_3(n)=\frac{p_3(n)}{h_3(n)} \ , \qquad \delta_3(n)=\frac{q_3(n)}{p_3(n)} 
\eeq
for a positive integer $n$ as in Eq. (\ref{ratiodef}). 
As $n$ increases, we find $\alpha_3(n)$ increases monotonically while $\beta_3(n)$, $\gamma_3(n)$, $\delta_3(n)$ decreases monotonically with the relation $\alpha_3(n) \le \beta_3(n) \le \gamma_3(n) \le \delta_3(n)$.
The values of these ratios for small $n$ are listed in Table \ref{tablesg3n}. Numerically, we find
\beq
\lim _{n \to \infty} \alpha_3(n) = \lim _{n \to \infty} \beta_3(n) = \lim _{n \to \infty} \gamma_3(n) = \lim _{n \to \infty} \delta_3(n) = 0.442256573677178603386...
\eeq

\begin{table}[htbp]
\caption{\label{tablesg3n} The first few values of $\alpha_3(n)$, $\beta_3(n)$, $\gamma_3(n)$, $\delta_3(n)$. The last digits given are rounded off.}
\begin{center}
\begin{tabular}{|c||r|r|r|r|}
\hline\hline 
$n$           &   1 &                 2 &                 3 & 4 \\ \hline\hline 
$\alpha_3(n)$ & 0.4 & 0.441065964287098 & 0.442255671189410 & 0.442256573676665 \\ \hline 
$\beta_3(n)$  & 0.5 & 0.443488188284886 & 0.442257510059261 & 0.442256573677711 \\ \hline 
$\gamma_3(n)$ & 0.5 & 0.445940594059406 & 0.442259349015113 & 0.442256573678758 \\ \hline 
$\delta_3(n)$ &   1 & 0.448490230905861 & 0.442261188057088 & 0.442256573679804 \\ \hline\hline 
\end{tabular}
\end{center}
\end{table}

By a similar argument as Lemma \ref{lemmasg2b}, the asymptotic growth constant for the number of independent sets on $SG_3(n)$ is bounded:
\beq
\frac{1}{2\times 4^m} \ln [f_3(m)] + \frac{1}{4^m} \ln [1+\alpha_3^2(m)] \le z_{SG_3} \le \frac{1}{2\times 4^m} \ln [f_3(m)] + \frac{1}{4^m} \ln [1+\delta_3^2(m)] \ ,
\label{zsg3}
\eeq
where $m$ is a positive integer. More than a hundred significant figures for $z_{SG_3}$ can be obtained when $m$ is equal to seven. We have the following proposition.

\bigskip

\begin{propo} \label{proposg3} The asymptotic growth constant for the number of independent sets on the three-dimensional Sierpinski gasket $SG_3(n)$ in the large $n$ limit is $z_{SG_3}=0.32859960572147955761...$.

\end{propo}

\section{The number of independent sets on $SG_4(n)$} 

In this section, we derive the asymptotic growth constant of independent sets on the four-dimensional Sierpinski gasket $SG_4(n)$. We use the following definitions.

\bigskip

\begin{defi} \label{defisg4} Consider the four-dimensional Sierpinski gasket $SG_4(n)$ at stage $n$. (i) Define $m_4(n) \equiv N_{IS}(SG_4(n))$ as the number of independent sets. (ii) Define $f_4(n)$ as the number of independent sets such that all five outmost vertices are not in the vertex subset. (iii) Define $g_4(n)$ as the number of independent sets such that one certain outmost vertex is in the vertex subset. (iv) Define $h_4(n)$ as the number of independent sets such that two certain outmost vertices are in the vertex subset. (v) Define $p_4(n)$ as the number of independent sets such that three certain outmost vertices are in the vertex subset. (vi) Define $q_4(n)$ as the number of independent sets such that four certain outmost vertices are in the vertex subset. (vi) Define $r_4(n)$ as the number of independent sets such that all five outmost vertices are in the vertex subset.
\end{defi}

\bigskip

\begin{figure}[htbp]
\unitlength 1.8mm 
\begin{picture}(90,9)
\put(2,0){\line(1,0){6}}
\put(2,0){\line(4,3){8}}
\put(2,0){\line(-1,3){2}}
\put(2,0){\line(1,3){3}}
\put(8,0){\line(-1,3){3}}
\put(8,0){\line(1,3){2}}
\put(8,0){\line(-4,3){8}}
\put(0,6){\line(1,0){10}}
\put(5,9){\line(5,-3){5}}
\put(5,9){\line(-5,-3){5}}
\put(2.5,0.5){\circle{1}}
\put(7.5,0.5){\circle{1}}
\put(1,5.7){\circle{1}}
\put(9,5.7){\circle{1}}
\put(5,8){\circle{1}}
\put(5,-2){\makebox(0,0){$f_4(n)$}}
\put(18,0){\line(1,0){6}}
\put(18,0){\line(4,3){8}}
\put(18,0){\line(-1,3){2}}
\put(18,0){\line(1,3){3}}
\put(24,0){\line(-1,3){3}}
\put(24,0){\line(1,3){2}}
\put(24,0){\line(-4,3){8}}
\put(16,6){\line(1,0){10}}
\put(21,9){\line(5,-3){5}}
\put(21,9){\line(-5,-3){5}}
\put(18.5,0.5){\circle{1}}
\put(23.5,0.5){\circle{1}}
\put(17,5.7){\circle{1}}
\put(25,5.7){\circle{1}}
\put(21,8){\circle*{1}}
\put(21,-2){\makebox(0,0){$g_4(n)$}}
\put(34,0){\line(1,0){6}}
\put(34,0){\line(4,3){8}}
\put(34,0){\line(-1,3){2}}
\put(34,0){\line(1,3){3}}
\put(40,0){\line(-1,3){3}}
\put(40,0){\line(1,3){2}}
\put(40,0){\line(-4,3){8}}
\put(32,6){\line(1,0){10}}
\put(37,9){\line(5,-3){5}}
\put(37,9){\line(-5,-3){5}}
\put(34.5,0.5){\circle*{1}}
\put(39.5,0.5){\circle*{1}}
\put(33,5.7){\circle{1}}
\put(41,5.7){\circle{1}}
\put(37,8){\circle{1}}
\put(37,-2){\makebox(0,0){$h_4(n)$}}
\put(50,0){\line(1,0){6}}
\put(50,0){\line(4,3){8}}
\put(50,0){\line(-1,3){2}}
\put(50,0){\line(1,3){3}}
\put(56,0){\line(-1,3){3}}
\put(56,0){\line(1,3){2}}
\put(56,0){\line(-4,3){8}}
\put(48,6){\line(1,0){10}}
\put(53,9){\line(5,-3){5}}
\put(53,9){\line(-5,-3){5}}
\put(50.5,0.5){\circle*{1}}
\put(55.5,0.5){\circle*{1}}
\put(49,5.7){\circle{1}}
\put(57,5.7){\circle{1}}
\put(53,8){\circle*{1}}
\put(53,-2){\makebox(0,0){$p_4(n)$}}
\put(66,0){\line(1,0){6}}
\put(66,0){\line(4,3){8}}
\put(66,0){\line(-1,3){2}}
\put(66,0){\line(1,3){3}}
\put(72,0){\line(-1,3){3}}
\put(72,0){\line(1,3){2}}
\put(72,0){\line(-4,3){8}}
\put(64,6){\line(1,0){10}}
\put(69,9){\line(5,-3){5}}
\put(69,9){\line(-5,-3){5}}
\put(66.5,0.5){\circle*{1}}
\put(71.5,0.5){\circle*{1}}
\put(65,5.7){\circle*{1}}
\put(73,5.7){\circle*{1}}
\put(69,8){\circle{1}}
\put(69,-2){\makebox(0,0){$q_4(n)$}}
\put(82,0){\line(1,0){6}}
\put(82,0){\line(4,3){8}}
\put(82,0){\line(-1,3){2}}
\put(82,0){\line(1,3){3}}
\put(88,0){\line(-1,3){3}}
\put(88,0){\line(1,3){2}}
\put(88,0){\line(-4,3){8}}
\put(80,6){\line(1,0){10}}
\put(85,9){\line(5,-3){5}}
\put(85,9){\line(-5,-3){5}}
\put(82.5,0.5){\circle*{1}}
\put(87.5,0.5){\circle*{1}}
\put(81,5.7){\circle*{1}}
\put(89,5.7){\circle*{1}}
\put(85,8){\circle*{1}}
\put(85,-2){\makebox(0,0){$r_4(n)$}}
\end{picture}

\vspace*{5mm}
\caption{\footnotesize{Illustration for the configurations $f_4(n)$, $g_4(n)$, $h_4(n)$, $p_4(n)$, $q_4(n)$ and $r_4(n)$. Only the five outmost vertices are shown explicitly, where a solid circle is in the vertex subset and an open circle is not.}} 
\label{fghpqrfig}
\end{figure}
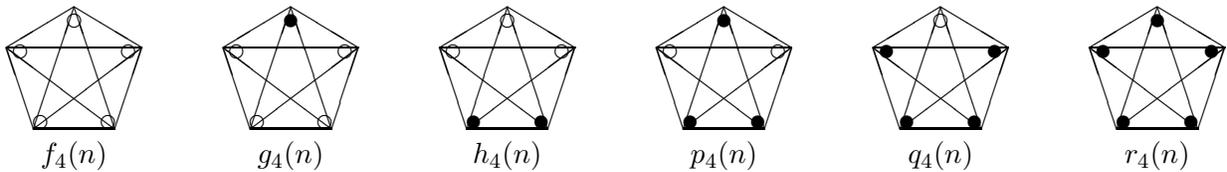

The quantities $f_4(n)$, $g_4(n)$, $h_4(n)$, $p_4(n)$, $q_4(n)$ and $r_4(n)$ are illustrated in Fig. \ref{fghpqrfig}, where only the outmost vertices are shown. There are ${5 \choose 1}=5$ equivalent $g_4(n)$, ${5 \choose 2}=10$ equivalent $h_4(n)$, ${5 \choose 3}=10$ equivalent $p_3(n)$, and ${5 \choose 4}=5$ equivalent $q_3(n)$. By definition,
\beq
m_4(n) = f_4(n)+5g_4(n)+10h_4(n)+10p_4(n)+5q_4(n)+r_4(n) \ .
\label{fsg4}
\eeq
The initial values at stage zero are $f_4(0)=1$, $g_4(0)=1$, $h_4(0)=0$, $p_4(0)=0$, $q_4(0)=0$, $r_4(0)=0$ and $m_4(0)=6$.

The recursion relations are lengthy and given in the appendix. Some values of $f_4(n)$, $g_4(n)$, $h_4(n)$, $p_4(n)$, $q_4(n)$, $r_4(n)$, $m_4(n)$ are listed in Table \ref{tablesg4}. These numbers grow exponentially, and do not have simple integer factorizations.

\begin{table}[htbp]
\caption{\label{tablesg4} The first few values of $f_4(n)$, $g_4(n)$, $h_4(n)$, $p_4(n)$, $q_4(n)$, $r_4(n)$, $m_4(n)$.}
\begin{center}
\begin{tabular}{|c||r|r|r|r|}
\hline\hline 
$n$      & 0 &   1 &           2 & 3 \\ \hline\hline 
$f_4(n)$ & 1 &  26 &  48,645,865 & 1,209,689,823,065,753,613,801,849,265,389,348,210,254 \\ \hline 
$g_4(n)$ & 1 &  10 &  19,499,025 & 485,275,031,765,121,996,003,377,748,244,728,141,942 \\ \hline 
$h_4(n)$ & 0 &   4 &   7,827,058 & 194,671,321,306,020,419,533,199,834,929,606,628,798 \\ \hline
$p_4(n)$ & 0 &   2 &   3,146,558 & 78,093,721,039,746,646,163,976,217,053,630,607,240 \\ \hline 
$q_4(n)$ & 0 &   1 &   1,266,948 & 31,327,833,873,772,900,771,790,623,812,192,536,505 \\ \hline
$r_4(n)$ & 0 &   1 &     510,980 & 12,567,379,442,065,248,794,102,222,711,306,394,841 \\ \hline
$m_4(n)$ & 6 & 142 & 262,722,870 & 6,532,921,954,159,964,003,443,553,868,217,630,357,710 \\ \hline\hline 
\end{tabular}
\end{center}
\end{table}

Define ratios
\beq
\alpha_4(n)=\frac{g_4(n)}{f_4(n)} \ , \quad \beta_4(n)=\frac{h_4(n)}{g_4(n)} \ , \quad \gamma_4(n)=\frac{p_4(n)}{h_4(n)} \ , \quad \delta_4(n)=\frac{q_4(n)}{p_4(n)} \ , \quad \eta_4(n)=\frac{r_4(n)}{q_4(n)} 
\eeq
for a positive integer $n$ as in Eq. (\ref{ratiodef}). As $n \ge 2$ increases, we find $\alpha_4(n)$ increases monotonically while $\beta_4(n)$, $\gamma_4(n)$, $\delta_4(n)$, $\eta_4(n)$ decreases monotonically with the relation $\alpha_4(n) \le \beta_4(n) \le \gamma_4(n) \le \delta_4(n) \le \eta_4(n)$.
The values of these ratios for small $n$ are listed in Table \ref{tablesg4n}. Numerically, we find
\beq
\lim _{n \to \infty} \alpha_4(n) = \lim _{n \to \infty} \beta_4(n) = \lim _{n \to \infty} \gamma_4(n) = \lim _{n \to \infty} \delta_4(n) = \lim _{n \to \infty} \eta_4(n) = 0.401156636030339443965...
\eeq

\begin{table}[htbp]
\caption{\label{tablesg4n} The first few values of $\alpha_4(n)$, $\beta_4(n)$, $\gamma_4(n)$, $\delta_4(n)$, $\eta_4(n)$. The last digits given are rounded off.}
\begin{center}
\begin{tabular}{|c||r|r|r|r|}
\hline\hline 
$n$           &                 1 &                 2 &                 3 & 4 \\ \hline\hline 
$\alpha_4(n)$ & 0.384615384615385 & 0.400836227292906 & 0.401156579572832 & 0.401156636030338 \\ \hline 
$\beta_4(n)$  &               0.4 & 0.401407660126596 & 0.401156681393497 & 0.401156636030341 \\ \hline 
$\gamma_4(n)$ &               0.5 & 0.402010308343186 & 0.401156783217105 & 0.401156636030344 \\ \hline 
$\delta_4(n)$ &               0.5 & 0.402645684586141 & 0.401156885043655 & 0.401156636030347 \\ \hline 
$\eta_4(n)$   &                 1 & 0.403315684621626 & 0.401156986873147 & 0.401156636030350 \\ \hline\hline 
\end{tabular}
\end{center}
\end{table}

By a similar argument as Lemma \ref{lemmasg2b}, the asymptotic growth constant for the number of independent sets on $SG_4(n)$ is bounded:
\beq
\frac{2}{5^{m+1}} \ln [f_4(m)] + \frac{1}{5^m} \ln [1+\alpha_4^2(m)] \le z_{SG_4} \le \frac{2}{5^{m+1}} \ln [f_4(m)] + \frac{1}{5^m} \ln [1+\eta_4^2(m)] \ ,
\label{zsg4}
\eeq
where $m$ is a positive integer. More than a hundred significant figures for $z_{SG_4}$ can be obtained when $m$ is equal to seven. We have the following proposition.

\bigskip

\begin{propo} \label{proposg4} The asymptotic growth constant for the number of independent sets on the four-dimensional Sierpinski gasket $SG_4(n)$ in the large $n$ limit is $z_{SG_4}=0.28916553234872775551...$.

\end{propo}

\section{Bounds of the asymptotic growth constants}

For the $d$-dimensional Sierpinski gasket $SG_d(n)$, we conjecture that similar upper and lower bounds for the asymptotic growth constant as in Lemma \ref{lemmasg2b} hold,
\beqs
& & \frac{2}{(d+1)^{m+1}} \ln [f_d(m)] + \frac{1}{(d+1)^m} \ln [1+\alpha_d^2(m)] \le z_{SG_d} \cr
& & \le \frac{2}{(d+1)^{m+1}} \ln [f_d(m)] + \frac{1}{(d+1)^m} \ln [1+\zeta_d^2(m)] 
\label{zsgd}
\eeqs
with a positive integer $m$, where the ratios are defined as
\beq
\alpha_d(n)=\frac{g_d(n)}{f_d(n)} \ , \qquad \zeta_d(n)=\frac{t_d(n)}{s_d(n)} \ , 
\eeq
for a positive integer $n$. $f_d(n)$ again is the number of independent sets such that all $d+1$ outmost vertices are not in the vertex subset, $g_d(n)$ is the number of independent sets such that one certain outmost vertex is in the vertex subset, $s_d(n)$ is the number of independent sets such that all but one certain outmost vertex are in the vertex subset, and $t_4(n)$ is the number of independent sets such that all $d+1$ outmost vertices are in the vertex subset.

Although the quantities in Eq. (\ref{zsgd}) for general $m$ is difficult to obtain, one can consider the simplest case $m=1$. Denote the upper and lowers bounds at $m=1$ as $\bar z_{SG_d}$ and $\underline z_{SG_d}$, respectively. Because $s_d(1)=t_d(1)=1$ and $g_d(1)=f_{d-1}(1)$, we have
\beqs
\bar z_{SG_d} & = & \frac{2}{(d+1)^2} \ln [f_d(1)] + \frac{1}{d+1} \ln (2) \ , \cr
\underline z_{SG_d} & = & \frac{2}{(d+1)^2} \ln [f_d(1)] + \frac{1}{d+1} \ln \Bigl[ 1 + \Bigl( \frac{f_{d-1}(1)}{f_d(1)} \Bigr)^2 \Bigr] \ ,
\label{zsgdn}
\eeqs
and the task reduces to the determination of $f_d(1)$. It is easy to see that $f_1(1)=2$ and assume the formal quantity $f_0(1)=1$, then $f_d(1)$ satisfies the recursion relation
\beq
f_d(1) = f_{d-1}(1) + df_{d-2}(1)
\label{fd}
\eeq
for $d \ge 2$. This relation can be understood as follows. The $d$-dimensional Sierpinski gasket $SG_d(1)$ at stage one is the juxtaposition of $d+1$ complete graphs $K_{d+1}$. For the enumeration of $f_d(1)$, consider one of the complete graphs. In the case that all $d$ interior vertices of the complete graph are not in the vertex subset, the number is the same as $g_d(1)=f_{d-1}(1)$, which is given as the first term on the right-hand-side of Eq. (\ref{fd}). In the case that one of the $d$ interior vertices of the complete graph is in the vertex subset, the number is given by $f_{d-2}(1)$, which gives the second term on the right-hand-side of Eq. (\ref{fd}). It follows that $f_d(1)$ is equal to the number of permutation involutions on $d+1$ elements. The values of $f_d(1)$, $\underline z_{SG_d}$, $\bar z_{SG_d}$ for small $d$ are listed in Table \ref{zsgdtable}. We notice that $\underline z_{SG_d}$ is closer to $z_{SG_d}$ compared with $\bar z_{SG_d}$, and serves as an approximation for $z_{SG_d}$. The fact that $z_{SG_d}$ decreases as $d$ increases is easy to understand because the degrees of the vertices of $SG_d(n)$ also increase.

\begin{table}
\caption{\label{zsgdtable} Numerical values of $\underline z_{SG_d}$, $\bar z_{SG_d}$, and some ratios of them to $z_{SG_d}$. The last digits given are rounded off.}
\begin{center}
\begin{tabular}{|r|r|c|c|c|c|c|}
\hline\hline 
$d$ & $f_d(1)$ & $\underline z_{SG_d}$ & $\bar z_{SG_d}$ & $z_{SG_d}$ & $\underline z_{SG_d}/z_{SG_d}$ & $\bar z_{SG_d}/z_{SG_d}$ \\ \hline\hline 
2   &     4 & 0.3824465974 & 0.5391144738 & 0.3843095344 & 0.9951525088 & 1.402813164 \\ \hline
3   &    10 & 0.3249281379 & 0.4611099318 & 0.3285996057 & 0.9888269257 & 1.403257715 \\ \hline
4   &    26 & 0.2882396119 & 0.3992771592 & 0.2891655323 & 0.9967979570 & 1.380790981 \\ \hline
5   &    76 & 0.2590427565 & 0.3561208268 & - & - & - \\ \hline
6   &   232 & 0.2368781125 & 0.3213368369 & - & - & - \\ \hline
7   &   764 & 0.2184809121 & 0.2940986410 & - & - & - \\ \hline
8   &  2620 & 0.2034116955 & 0.2713602941 & - & - & - \\ \hline
9   &  9496 & 0.1905090814 & 0.2524872368 & - & - & - \\ \hline
10  & 35696 & 0.1794854089 & 0.2362827010 & - & - & - \\ \hline\hline
\end{tabular}
\end{center}
\end{table}

\section*{Acknowledgments} 
The research of S.C.C. was partially supported by the NSC grant NSC-97-2112-M-006-007-MY3 and NSC-99-2119-M-002-001. The research of L.C.C. was partially supported by the NSC grant NSC-99-2115-M-030-004-MY3. The research of W.Y. was partially supported by the NSFC grant 10771086.

\appendix

\section{Recursion relations for $SG_{2,3}(n)$}

We give the recursion relations for the generalized two-dimensional Sierpinski gasket $SG_{2,3}(n)$ here. Since the subscript is $d=2,b=3$ for all the quantities throughout this section, we will use the simplified notation $f_{n+1}$ to denote $f_{2,3}(n+1)$ and similar notations for other quantities. For any non-negative integer $n$, we have
\beqs
f_{n+1} & = & f_n^6 + 6f_n^4g_n^2 + 9f_n^2g_n^4 + 6f_n^3g_n^2h_n + 2g_n^6 + 12f_ng_n^4h_n + 6f_n^2g_n^2h_n^2 + 9g_n^4h_n^2 + 6f_ng_n^2h_n^3 \cr
& & + 6g_n^2h_n^4 + h_n^6 + f_n^3g_n^3 + 6f_n^2g_n^3h_n + 9f_ng_n^3h_n^2 + 3f_n^2g_nh_n^3 + 3f_ng_n^4p_n + 2g_n^3h_n^3 \cr
& & + 6f_ng_n^2h_n^2p_n + 6g_n^4h_np_n + 6f_ng_nh_n^4 + 3f_nh_n^4p_n + 9g_n^2h_n^3p_n + 3g_n^3h_np_n^2 + 6g_nh_n^3p_n^2 \cr
& & + h_n^3p_n^3 \ , 
\eeqs
\beqs
g_{n+1} & = & f_n^5g_n + 2f_n^4g_nh_n + 4f_n^3g_n^3 + 3f_ng_n^5 + 9f_n^2g_n^3h_n + 2f_n^3g_nh_n^2 + f_n^3g_n^2p_n + 4g_n^5h_n \cr
& & + 10f_ng_n^3h_n^2 + 2f_n^2g_nh_n^3 + 2f_ng_n^4p_n + 2f_n^2g_n^2h_np_n + 7g_n^3h_n^3 + 3f_ng_n^2h_n^2p_n + 3g_n^4h_np_n \cr
& & + 2f_ng_nh_n^4 + 2g_nh_n^5 + 4g_n^2h_n^3p_n + h_n^5p_n + f_n^2g_n^4 + 4f_ng_n^4h_n + 2f_n^2g_n^2h_n^2 + 3g_n^4h_n^2 \cr
& & + 8f_ng_n^2h_n^3 + 2f_ng_n^3h_np_n + f_n^2g_nh_n^2p_n + g_n^5p_n + 4g_n^2h_n^4 + 4f_ng_nh_n^3p_n + 8g_n^3h_n^2p_n \cr
& & + 2f_ng_n^2h_np_n^2 + 2f_nh_n^5 + 7g_nh_n^4p_n + 5g_n^2h_n^2p_n^2 + 2f_nh_n^3p_n^2 + g_n^3p_n^3 + 2h_n^4p_n^2 + 4g_nh_n^2p_n^3 \cr
& & + h_n^2p_n^4 \ , 
\eeqs
\beqs
h_{n+1} & = & f_n^4g_n^2 + 2f_n^2g_n^4 + 4f_n^3g_n^2h_n + 7f_ng_n^4h_n + 5f_n^2g_n^2h_n^2 + 2f_n^2g_n^3p_n + f_n^3h_n^3 + 4g_n^4h_n^2 \cr
& & + 8f_ng_n^2h_n^3 + 4f_ng_n^3h_np_n + 2f_n^2g_nh_n^2p_n + 2g_n^5p_n + 3g_n^2h_n^4 + 2f_ng_nh_n^3p_n + 8g_n^3h_n^2p_n \cr
& & + f_ng_n^2h_np_n^2 + f_nh_n^5 + 4g_nh_n^4p_n + 2g_n^2h_n^2p_n^2 + h_n^4p_n^2 + f_ng_n^5 + 2g_n^5h_n + 4f_ng_n^3h_n^2 \cr
& & + 7g_n^3h_n^3 + 3f_ng_n^2h_n^2p_n + 2g_n^4h_np_n + 3f_ng_nh_n^4 + 4g_nh_n^5 + 2f_nh_n^4p_n + 10g_n^2h_n^3p_n \cr
& & + 2g_n^3h_np_n^2 + 2f_ng_nh_n^2p_n^2 + 9g_nh_n^3p_n^2 + 3h_n^5p_n + 2g_n^2h_np_n^3 + f_nh_n^2p_n^3 + 4h_n^3p_n^3 + 2g_nh_np_n^4 \cr
& & + h_np_n^5 \ , 
\eeqs
\beqs
p_{n+1} & = & f_n^3g_n^3 + 6f_n^2g_n^3h_n + 9f_ng_n^3h_n^2 + 3f_n^2g_nh_n^3 + 3f_ng_n^4p_n + 2g_n^3h_n^3 + 6f_ng_n^2h_n^2p_n + 6g_n^4h_np_n \cr
& & + 6f_ng_nh_n^4 + 3f_nh_n^4p_n + 9g_n^2h_n^3p_n + 3g_n^3h_np_n^2 + 6g_nh_n^3p_n^2 + h_n^3p_n^3 + g_n^6 + 6g_n^4h_n^2 \cr
& & + 9g_n^2h_n^4 + 6g_n^3h_n^2p_n + 2h_n^6 + 12g_nh_n^4p_n + 6g_n^2h_n^2p_n^2 + 9h_n^4p_n^2 + 6g_nh_n^2p_n^3 + 6h_n^2p_n^4 + p_n^6 \ . \cr
& & 
\eeqs
There are always $128=2^7$ terms (counting multiplicity) in these equations.

\section{Recursion relations for $SG_3(n)$}

We give the recursion relations for the three-dimensional Sierpinski gasket $SG_3(n)$ here. Since the subscript is $d=3$ for all the quantities throughout this section, we will use the simplified notation $f_{n+1}$ to denote $f_3(n+1)$ and similar notations for other quantities. For any non-negative integer $n$, we have
\beqs
f_{n+1} & = & f_n^4 + 6f_n^2g_n^2 + 12f_ng_n^2h_n + 3g_n^4 + 4f_nh_n^3 + 12g_n^2h_n^2 + 4g_n^3p_n + 3h_n^4 + 12g_nh_n^2p_n \cr
& & + 6h_n^2p_n^2 + p_n^4 \ , 
\eeqs
\beqs
g_{n+1} & = & f_n^3g_n + 3f_n^2g_nh_n + 3f_ng_n^3 + 6f_ng_nh_n^2 + 6g_n^3h_n + 3f_ng_n^2p_n + 7g_nh_n^3 + 3f_nh_n^2p_n \cr
& & + 9g_n^2h_np_n + g_n^3q_n + 6h_n^3p_n + 6g_nh_np_n^2 + 3g_nh_n^2q_n + 3h_np_n^3 + 3h_n^2p_nq_n + p_n^3q_n \ , 
\eeqs
\beqs
h_{n+1} & = & f_n^2g_n^2 + 4f_ng_n^2h_n + f_n^2h_n^2 + g_n^4 + 2f_nh_n^3 + 7g_n^2h_n^2 + 2g_n^3p_n + 4f_ng_nh_np_n + 2h_n^4 \cr
& & + 12g_nh_n^2p_n + 2g_n^2p_n^2 + 2f_nh_np_n^2 + 2g_n^2h_nq_n + 7h_n^2p_n^2 + 2g_np_n^3 + 2h_n^3q_n + 4g_nh_np_nq_n \cr
& & + p_n^4 + 4h_np_n^2q_n + h_n^2q_n^2 + p_n^2q_n^2 \ , 
\eeqs
\beqs
p_{n+1} & = & f_ng_n^3 + 3f_ng_nh_n^2 + 3g_n^3h_n + 6g_nh_n^3 + 3f_nh_n^2p_n + 6g_n^2h_np_n + 7h_n^3p_n + 9g_nh_np_n^2 + f_np_n^3 \cr
& & + 3g_nh_n^2q_n + 6h_np_n^3 + 6h_n^2p_nq_n + 3g_np_n^2q_n + 3p_n^3q_n + 3h_np_nq_n^2 + p_nq_n^3 \ , 
\eeqs
\beqs
q_{n+1} & = & g_n^4 + 6g_n^2h_n^2 + 3h_n^4 + 12g_nh_n^2p_n + 12h_n^2p_n^2 + 4g_np_n^3 + 4h_n^3q_n + 3p_n^4 + 12h_np_n^2q_n \cr
& & + 6p_n^2q_n^2 + q_n^4 \ .
\eeqs
There are always $64=2^6$ terms (counting multiplicity) in these equations.

\section{Recursion relations for $SG_4(n)$}

We give the recursion relations for the four-dimensional Sierpinski gasket $SG_4(n)$ here. Since the subscript is $d=4$ for all the quantities throughout this section, we will use the simplified notation $f_{n+1}$ to denote $f_4(n+1)$ and similar notations for other quantities. For any non-negative integer $n$, we have
\beqs
f_{n+1} & = & f_n^5 + 10f_n^3g_n^2 + 30f_n^2g_n^2h_n + 15f_ng_n^4 + 10f_n^2h_n^3 + 60f_ng_n^2h_n^2 + 20f_ng_n^3p_n + 30g_n^4h_n \cr
& & + 15f_nh_n^4 + 60f_ng_nh_n^2p_n + 70g_n^2h_n^3 + 60g_n^3h_np_n + 5g_n^4q_n + 30f_nh_n^2p_n^2 + 12h_n^5 \cr
& & + 120g_nh_n^3p_n + 60g_n^2h_np_n^2 + 30g_n^2h_n^2q_n + 5f_np_n^4 + 70h_n^3p_n^2 + 60g_nh_np_n^3 + 15h_n^4q_n \cr
& & + 60g_nh_n^2p_nq_n + 30h_np_n^4 + 60h_n^2p_n^2q_n + 10h_n^3q_n^2 + 20g_np_n^3q_n + 30h_np_n^2q_n^2 + 15p_n^4q_n \cr
& & + 10p_n^2q_n^3 + q_n^5 \ , 
\eeqs
\beqs
g_{n+1} & = & f_n^4g_n + 4f_n^3g_nh_n + 6f_n^2g_n^3 + 12f_n^2g_nh_n^2 + 24f_ng_n^3h_n + 6f_n^2g_n^2p_n + 3g_n^5 + 28f_ng_nh_n^3 \cr
& & + 6f_n^2h_n^2p_n + 36f_ng_n^2h_np_n + 4f_ng_n^3q_n + 36g_n^3h_n^2 + 10g_n^4p_n + 24f_nh_n^3p_n + 24f_ng_nh_np_n^2 \cr
& & + 12f_ng_nh_n^2q_n + 31g_nh_n^4 + 90g_n^2h_n^2p_n + 12g_n^3p_n^2 + 16g_n^3h_nq_n + g_n^4r_n + 12f_nh_np_n^3 \cr
& & + 12f_nh_n^2p_nq_n + 36h_n^4p_n + 102g_nh_n^2p_n^2 + 12g_n^2p_n^3 + 36g_n^2h_np_nq_n + 36g_nh_n^3q_n + 6g_n^2h_n^2r_n \cr
& & + 4f_np_n^3q_n + 54h_n^2p_n^3 + 52h_n^3p_nq_n + 60g_nh_np_n^2q_n + 3h_n^4r_n + 12g_nh_n^2p_nr_n + 13g_np_n^4 \cr
& & + 12g_nh_n^2q_n^2 + 52h_np_n^3q_n + 30h_n^2p_nq_n^2 + 12h_n^2p_n^2r_n + 6p_n^5 + 4h_n^3q_nr_n + 12g_np_n^2q_n^2 \cr
& & + 4g_np_n^3r_n + 12h_np_nq_n^3 + 12h_np_n^2q_nr_n + 18p_n^3q_n^2 + 3p_n^4r_n + 4p_nq_n^4 + 6p_n^2q_n^2r_n + q_n^4r_n \ , \cr
& & 
\eeqs
\beqs
h_{n+1} & = & f_n^3g_n^2 + 6f_n^2g_n^2h_n + f_n^3h_n^2 + 3f_ng_n^4 + 3f_n^2h_n^3 + 21f_ng_n^2h_n^2 + 6f_ng_n^3p_n + 6f_n^2g_nh_np_n \cr
& & + 9g_n^4h_n + 6f_nh_n^4 + 36f_ng_nh_n^2p_n + 6f_ng_n^2p_n^2 + 3f_n^2h_np_n^2 + 31g_n^2h_n^3 + 6f_ng_n^2h_nq_n \cr
& & + 30g_n^3h_np_n + 2g_n^4q_n + 21f_nh_n^2p_n^2 + 6f_ng_np_n^3 + 6f_nh_n^3q_n + 12f_ng_nh_np_nq_n + 7h_n^5 \cr
& & + 72g_nh_n^3p_n + 51g_n^2h_np_n^2 + 27g_n^2h_n^2q_n + 6g_n^3p_nq_n + 2g_n^3h_nr_n + 3f_np_n^4 + 12f_nh_np_n^2q_n \cr
& & + 3f_nh_n^2q_n^2 + 54h_n^3p_n^2 + 54g_nh_np_n^3 + 15h_n^4q_n + 78g_nh_n^2p_nq_n + 6g_n^2h_nq_n^2 + 15g_n^2p_n^2q_n \cr
& & + 6g_nh_n^3r_n + 6g_n^2h_np_nr_n + 3f_np_n^2q_n^2 + 27h_np_n^4 + 81h_n^2p_n^2q_n + 30g_nh_np_nq_n^2 + 13h_n^3q_n^2 \cr
& & + 26g_np_n^3q_n + 12h_n^3p_nr_n + 12g_nh_np_n^2r_n + 6g_nh_n^2q_nr_n + 51h_np_n^2q_n^2 + 18p_n^4q_n + 6h_n^2q_n^3 \cr
& & + 6g_np_nq_n^3 + 14h_np_n^3r_n + 18h_n^2p_nq_nr_n + h_n^3r_n^2 + 6g_np_n^2q_nr_n + 3h_nq_n^4 + 15p_n^2q_n^3 \cr
& & + 12h_np_nq_n^2r_n + 3h_np_n^2r_n^2 + 12p_n^3q_nr_n + q_n^5 + 6p_nq_n^3r_n + 3p_n^2q_nr_n^2 + q_n^3r_n^2 \ , 
\eeqs
\beqs
p_{n+1} & = & f_n^2g_n^3 + 3f_n^2g_nh_n^2 + 6f_ng_n^3h_n + g_n^5 + 12f_ng_nh_n^3 + 3f_n^2h_n^2p_n + 12f_ng_n^2h_np_n + 15g_n^3h_n^2 \cr
& & + 3g_n^4p_n + 14f_nh_n^3p_n + 18f_ng_nh_np_n^2 + f_n^2p_n^3 + 6f_ng_nh_n^2q_n + 18g_nh_n^4 + 51g_n^2h_n^2p_n \cr
& & + 6g_n^3p_n^2 + 6g_n^3h_nq_n + 12f_nh_np_n^3 + 12f_nh_n^2p_nq_n + 6f_ng_np_n^2q_n + 27h_n^4p_n + 81g_nh_n^2p_n^2 \cr
& & + 13g_n^2p_n^3 + 30g_n^2h_np_nq_n + 26g_nh_n^3q_n + 3g_n^2h_n^2r_n + 6f_np_n^3q_n + 6f_nh_np_nq_n^2 + 54h_n^2p_n^3 \cr
& & + 54h_n^3p_nq_n + 78g_nh_np_n^2q_n + 6g_n^2p_nq_n^2 + 3h_n^4r_n + 12g_nh_n^2p_nr_n + 15g_np_n^4 + 15g_nh_n^2q_n^2 \cr
& & + 3g_n^2p_n^2r_n + 2f_np_nq_n^3 + 72h_np_n^3q_n + 51h_n^2p_nq_n^2 + 21h_n^2p_n^2r_n + 7p_n^5 + 6h_n^3q_nr_n \cr
& & + 27g_np_n^2q_n^2 + 6g_nh_nq_n^3 + 12g_nh_np_nq_nr_n + 6g_np_n^3r_n + 30h_np_nq_n^3 + 36h_np_n^2q_nr_n \cr
& & + 31p_n^3q_n^2 + 6h_n^2q_n^2r_n + 6p_n^4r_n + 6g_np_nq_n^2r_n + 2g_nq_n^4 + 3h_n^2p_nr_n^2 + 6h_nq_n^3r_n + 9p_nq_n^4 \cr
& & + 21p_n^2q_n^2r_n + 6h_np_nq_nr_n^2 + 3p_n^3r_n^2 + 3q_n^4r_n + 6p_nq_n^2r_n^2 + p_n^2r_n^3 + q_n^2r_n^3 \ , 
\eeqs
\beqs
q_{n+1} & = & f_ng_n^4 + 6f_ng_n^2h_n^2 + 4g_n^4h_n + 3f_nh_n^4 + 12f_ng_nh_n^2p_n + 18g_n^2h_n^3 + 12g_n^3h_np_n + 12f_nh_n^2p_n^2 \cr
& & + 4f_ng_np_n^3 + 4f_nh_n^3q_n + 6h_n^5 + 52g_nh_n^3p_n + 30g_n^2h_np_n^2 + 12g_n^2h_n^2q_n + 3f_np_n^4 \cr
& & + 12f_nh_np_n^2q_n + 54h_n^3p_n^2 + 52g_nh_np_n^3 + 13h_n^4q_n + 60g_nh_n^2p_nq_n + 12g_n^2p_n^2q_n + 4g_nh_n^3r_n \cr
& & + 6f_np_n^2q_n^2 + 36h_np_n^4 + 102h_n^2p_n^2q_n + 36g_nh_np_nq_n^2 + 12h_n^3q_n^2 + 36g_np_n^3q_n + 12h_n^3p_nr_n \cr
& & + 12g_nh_np_n^2r_n + f_nq_n^4 + 90h_np_n^2q_n^2 + 31p_n^4q_n + 12h_n^2q_n^3 + 16g_np_nq_n^3 + 24h_np_n^3r_n \cr
& & + 24h_n^2p_nq_nr_n + 12g_np_n^2q_nr_n + 10h_nq_n^4 + 36p_n^2q_n^3 + 36h_np_nq_n^2r_n + 6h_np_n^2r_n^2 + 28p_n^3q_nr_n \cr
& & + 4g_nq_n^3r_n + 3q_n^5 + 6h_nq_n^2r_n^2 + 24p_nq_n^3r_n + 12p_n^2q_nr_n^2 + 6q_n^3r_n^2 + 4p_nq_nr_n^3 + q_nr_n^4 \ ,
\eeqs
\beqs
r_{n+1} & = & g_n^5 + 10g_n^3h_n^2 + 15g_nh_n^4 + 30g_n^2h_n^2p_n + 30h_n^4p_n + 60g_nh_n^2p_n^2 + 10g_n^2p_n^3 + 20g_nh_n^3q_n \cr
& & + 70h_n^2p_n^3 + 60h_n^3p_nq_n + 60g_nh_np_n^2q_n + 5h_n^4r_n + 15g_np_n^4 + 120h_np_n^3q_n + 60h_n^2p_nq_n^2 \cr
& & + 30h_n^2p_n^2r_n + 12p_n^5 + 30g_np_n^2q_n^2 + 60h_np_nq_n^3 + 60h_np_n^2q_nr_n + 70p_n^3q_n^2 + 15p_n^4r_n \cr
& & + 5g_nq_n^4 + 20h_nq_n^3r_n + 30p_nq_n^4 + 60p_n^2q_n^2r_n + 10p_n^3r_n^2 + 15q_n^4r_n + 30p_nq_n^2r_n^2 + 10q_n^2r_n^3 \cr
& & + r_n^5 \ .
\eeqs
There are always $1024=2^{10}$ terms (counting multiplicity) in these equations.

\end{document}